# Relaxed Binaural LCMV Beamforming

Andreas I. Koutrouvelis, Richard C. Hendriks, Richard Heusdens and Jesper Jensen

*Abstract*—In this paper we propose a new binaural beamforming technique which can be seen as a relaxation of the linearly constrained minimum variance (LCMV) framework. The proposed method can achieve simultaneous noise reduction and exact binaural cue preservation of the target source, similar to the binaural minimum variance distortionless response (BMVDR) method. However, unlike BMVDR, the proposed method is also able to preserve the binaural cues of multiple interferers to a certain predefined accuracy. Specifically, it is able to control the trade-off between noise reduction and binaural cue preservation of the interferers by using a separate trade-off parameter per-interferer. Moreover, we provide a robust way of selecting these trade-off parameters in such a way that the preservation accuracy for the binaural cues of the interferers is always better than the corresponding ones of the BMVDR. The relaxation of the constraints in the proposed method achieves approximate binaural cue preservation of more interferers than other previously presented LCMV-based binaural beamforming methods that use strict equality constraints.

## I. INTRODUCTION

COMPARED to normal-hearing people, hearing-impaired people generally have more difficulties in understanding a target talker in complex acoustic environments with multiple interfering sources. To improve speech quality and intelligibility, single-microphone (see e.g. [1] for an overview) or multi-microphone noise reduction algorithms (see e.g., [2] for an overview) can be used. While the former are mostly effective in improving speech quality and reducing listening effort, the latter are also effective in improving speech intelligibility [3]. Examples of multi-microphone noise reduction algorithms include the multi-channel Wiener filter (MWF) [4], [5], the minimum variance distortionless response (MVDR) beamformer [6], [7], or, its generalization, the linearly constrained minimum variance (LCMV) beamformer [7], [8].

Traditionally, hearing aids have been fitted *bilaterally*, i.e., the user wears a hearing aid on each ear, and the hearing aids are operating essentially independently of each other. As such, the noise reduction algorithm in each hearing aid estimates the signal of interest using only the recordings of the microphones from that specific hearing aid [9]. Such a setup with an independent multi-microphone algorithm per ear may severely distort the binaural cues since phase and magnitude relations of the sources reaching the two ears are modified [10]. This is harmful for the naturalness of the total sound field as received by the hearing-aid user. Ideally, all sound sources (including the undesired ones) that are present after processing should still sound as if originating from the original direction. This does not only lead to a more natural perception of the acoustic environment, but can also lead to an improved intelligibility of a target speaker in certain cases; more specifically, in spatial unmasking experiments [11] it has been shown that a target speaker in a noisy background is significantly easier to understand when the noise sources are separated in space from the speaker, as compared to the situation where speaker and noise sources are co-located.

*Binaural* hearing aids are able to wirelessly exchange microphone signals between hearing aids. This facilitates the use of multi-microphone noise reduction algorithms which combine all microphone recordings from both hearing aids, hence allowing the usage of more microphone recordings than with the bilateral noise reduction. As such, the increased number of microphone recordings can potentially lead to better noise suppression and, thus, to a higher speech intelligibility. Moreover, by introducing proper constraints on the beamformer coefficients, binaural cue preservation of the sources can be achieved.

The LCMV algorithm [7], [8] minimizes the output power of the noise under multiple linear equality constraints. One of these equality constraints is typically used to guarantee that the target source remains undistorted with respect to a certain reference location or microphone. The remaining constraints can be used as additional degrees of freedom in designing the final filter. For example, it can be used to steer nulls in the directions of the interferers [7], [12], or to broaden the beam towards the target source in order to avoid pointing error problems, also known as steering vector mismatches [13], [14]. A special case of the LCMV method is the minimum variance distortionless response (MVDR) beamformer, which only uses the distortionless constraint of the target source [6], [7].

An alternative multi-microphone noise reduction method is the MWF which leads to the minimum mean square error (MMSE) estimate of the target source if the estimator is constrained to be linear, or, the target source and the noise are assumed to be jointly Gaussian distributed [15]. However, in [16]–[18], it was demonstrated that speech signals in time and frequency domains tend to be super-Gaussian distributed rather than Gaussian distributed. Thus, the MWF is generally not MMSE optimal. The MWF does not include a distortionless constraint for the target source and, thus, it generally introduces speech distortion in the output [4]. Several generalizations of the MWF have been proposed, among which the speech distortion weighted MWF (SDW-MWF) [5], which introduces a parameter in the minimization procedure to control the trade-off between speech distortion and noise reduction. A well-known property of the MWF is the fact that it can be decomposed into an MVDR beamformer and a single-channel Wiener filter as a post-processor [19]. Notice that this holds in general, also if the filter is not constrained to be linear and the target source is not assumed to be Gaussian

This work was supported by the Oticon Foundation and the Dutch Technology Foundation STW.

distributed [15].

There are several binaural multi-microphone noise reduction methods known from the literature. These can be devided into two main categories [20]: a) methods based on the linearly constrained minimum variance (LCMV) framework and b) methods based on the multi-channel Wiener filter (MWF).

The binaural version of the SDW-MWF (BSDW-MWF) [21], [22] preserves the binaural cues of the target. However, it was theoretically proven that the binaural cues of the interferers collapse on the binaural cues of the target source [23] (i.e., after processing the binaural cues of the interferers become identical to the binaural cues of the target source). In [22], a variation of the BSDW-MWF (called BSDW-MWF-N) was proposed which tries to partially preserve the binaural cues of the interferers. This method inserts a portion of the unprocessed noisy signal at the reference microphones to the coresponding BSDW-MWF enhanced signals. The larger the portion of the unprocessed noisy signals, the lower the noise reduction, but the better the preservation of binaural cues of the interferers and vice versa. As such, this solution exhibits a trade-off between the preservation of binaural cues and the amount of noise reduction. In [24], a subjective evaluation of BSDW-MWF and BSDW-MWF-N shows that for a moderate input SNR indeed the subjects localized the processed interferer correctly with BSDW-MWF-N and incorrectly with BSDW-MWF. However, for a small input SNR the processed interferer was also localized correctly for BSDW-MWF. This is mainly due to the inaccurate estimates of the cross power spectral density (CPSD) matrix of the target, and due to masking effects when the processed target and processed interferer are represented to the subjects simultaneously [24]. In [25], two other variations of the BSDW-MWF were proposed. The first one is capable of preserving the binaural cues of the target and completely cancel one interferer. The second one is capable of accurately preserving the binaural cues of only one interferer, while distorting the binaural cues of the target.

Similarly to SDW-MWF, the BSDW-MWF can be decomposed into the binaural MVDR (BMVDR) beamformer and a single-channel Wiener filter [25]. The BMVDR can preserve the binaural cues of the target source, but the binaural cues of the interferers collapse to the binaural cues of the target source. In [26], [27], the binaural linearly constrained minimum variance (BLCMV) algorithm was proposed, which achieves simultaneous noise reduction and binaural cue preservation of the target source and multiple interferers. Unlike the BMVDR, the BLCMV uses two additional linear constraints per interferer to preserve its binaural cues. A fixed interference rejection parameter is used in combination with these constraints to control the amount of noise reduction. The BLCMV is thus capable of controlling the amount of noise reduction using two constraints per interferer. However, in hearing-aid systems with a rather limited number of microphones, the degrees of freedom for noise reduction are exhausted quickly when increasing the number of interferers. This makes the BLCMV less suitable for this application.

In [28], a similar method to BLCMV, called optimal BLCMV (OBLCMV), was proposed which is able to achieve simultaneous noise reduction and binaural cue preservation of the target source and only one interferer. Unlike the BLCMV, the OBLCMV uses an optimal interference rejection parameter with respect to the binaural output SNR. In [29], [30] two independent works proposed the same LCMV-based method (we call it joint BLCMV (JBLCMV)) as an alternative to the BLCMV, which preserves the binaural cues of the target source and *more* than twice the number of interferers compared to the BLCMV [29]. Unlike the BLCMV, the JBLCMV requires only one linear constraint per interferer and, as a result, it has more degrees of freedom left for noise reduction. The linear constraints for the preservation of the binaural cues of the interferers have the same form as the linear constraint used in [25]. However, unlike the method in [25], the JBLCMV can preserve the binaural cues of a limited number of interferers and does not distort the binaural cues of the target source.

In this paper, we present an iterative, relaxed binaural LCMV beamforming method. Similar to the other binaural LCMV-based approaches, the proposed method strictly preserves the binaural cues of the target source. However, the proposed method is flexible to control the accuracy of binaural cue preservation of the interferers and, therefore, trade-off against additional noise reduction. This is achieved by using inequality constraints instead of the commonly used equality constraints. The task of each inequality constraint is the (approximate) preservation of the binaural cues of a single interferer in a controlled way. The proposed method is flexible to select a different value for the trade-off parameter of each interferer according to importance. The BMVDR and the JBLCMV can be seen as two extreme cases of the proposed method. On one hand, the BMVDR can achieve the best possible overall noise suppression compared to all the other aforementioned binaural LCMV-based methods, but causes full collapse of the binaural cues of the interferers towards the binaural cues of the target source. On the other hand, the JBLCMV can achieve the preservation of the maximum possible number of interferers compared to the other aforementioned binaural LCMV-based methods, but at the expense of less noise suppression. Unlike the JBLCMV and the BMVDR, the proposed method, is flexible to control the amount of noise suppression and binaural cue preservation according to the needs of the user. The relaxations used in the proposed method allow the usage of a substantially larger number of constraints for the approximate preservation of more interferers compared to all the other binaural LCMV-based methods including JBLCMV.

The remainder of this paper is organized as follows. In Section II, the signal model and the notation are presented. In Section III the key idea of the binaural beamforming is explained and several existing binaural LCMV-based algorithms are summarized. In Sections IV and V, a novel nonconvex binaural beamforming problem and its iterative convex approximation are presented, respectively. In Section VI, the evaluation of the proposed algorithm is provided. Finally, in Section VII, we draw some conclusions.

## II. SIGNAL MODEL AND NOTATION

Assume for convenience that each of the two hearing aids consists of $M/2$ microphones, where $M$ is an even number.

Therefore, the microphone array consists of $M$ microphones in total. The multi-microphone noise reduction methods considered in this paper operate in the frequency domain on a frame-by-frame basis. Let $l$ denote the frame index and $k$ the frequency-bin index. Assume that there is only one target source and there are $r$ interferers. The $k$-th frequency coefficient of the $l$-th frame of the $j$-th microphone noisy signal, $y_j(k, l)$, $j = 1, \cdots, M$, is given by

$$y_j(k,l) = \underbrace{a_j(k,l)s(k,l)}_{x_j(k,l)} + \sum_{i=1}^{r} \underbrace{b_{ij}(k,l)u_i(k,l)}_{n_{ij}(k,l)} + v_j(k,l),$$ (1)

where

- $s(k, l)$ denotes the target signal at the source location.
- $u_i(k, l)$, is the $i$-th interfering signal at the source location.
- $a_j(k, l)$ is the acoustic transfer function (ATF) of the target signal with respect to the $j$-th microphone.
- $b_{ij}(k, l)$ is the ATF of the $i$-th interfering signal with respect to the $j$-th microphone.
- $x_j(k, l)$ is the received target signal at the $j$-th microphone.
- $n_{ij}(k, l)$ is the $i$-th received interfering signal at the $j$-th microphone.
- $v_j(k, l)$ is additive noise at the $j$-th microphone.

Here we use in the signal model the ATFs for notational convinience. However, note that the ATFs can be replaced with relative acoustic transfer functions (RATFs) which can often be identified easier than the ATFs [12], [20].

In the remainder of the paper, the frequency and frame indices are neglected to simplify the notation. Using vector notation, Eq. (1) can be written as

$$\mathbf{y} = \mathbf{x} + \sum_{i=1}^{r} \mathbf{n}_i + \mathbf{v},$$ (2)

where $\mathbf{y} \in \mathbb{C}^{M \times 1}$, $\mathbf{x} \in \mathbb{C}^{M \times 1}$, $\mathbf{n}_i \in \mathbb{C}^{M \times 1}$ and $\mathbf{v} \in \mathbb{C}^{M \times 1}$ are the stacked vectors of the $y_j$, $x_j, n_{ij}, v_j$ (for $j = 1, \cdots, M$) components, respectively. Moreover, $\mathbf{x} = \mathbf{a}s$ and $\mathbf{n}_i = \mathbf{b}_i u_i$, where $\mathbf{a} \in \mathbb{C}^{M \times 1}$ and $\mathbf{b}_i \in \mathbb{C}^{M \times 1}$ are the stacked vectors of the $a_j$ and $b_{ij}$ (for $j = 1, \cdots, M$) components, respectively.

Assuming that all sources and the additive noise are mutually uncorrelated, the CPSD matrix $\mathbf{P_y} = E[\mathbf{y}\mathbf{y}^H]$ of $\mathbf{y}$ is given by

$$\mathbf{P_y} = \mathbf{P_x} + \underbrace{\sum_{i=1}^{r} \mathbf{P_{n_i}} + \mathbf{P_v}}_{\mathbf{P}},$$ (3)

where

- $\mathbf{P_x} = E[\mathbf{x}\mathbf{x}^H] = p_s \mathbf{a}\mathbf{a}^H \in \mathbb{C}^{M \times M}$ is the CPSD matrix of $\mathbf{x}$, with $p_s = E[|s|^2]$ the power spectral density (PSD) of $s$.
- $\mathbf{P_{n_i}} = E[\mathbf{n}_i \mathbf{n}_i^H] = p_{u_i} \mathbf{b}_i \mathbf{b}_i^H \in \mathbb{C}^{M \times M}$ is the CPSD matrix of $\mathbf{n}_i$, with $p_{u_i} = E[|u_i|^2]$ the PSD of $u_i$.
- $\mathbf{P_v} = E[\mathbf{v}\mathbf{v}^H] \in \mathbb{C}^{M \times M}$ is the CPSD matrix of $\mathbf{v}$.

- $\mathbf{P} = \sum_{i=1}^{r} \mathbf{P_{n_i}} + \mathbf{P_v}$ is the total CPSD matrix of all disturbances.

## III. Binaural Beamforming

Binaural multi-microphone noise reduction methods aim at the simultaneous noise reduction and binaural cue preservation of the sources. In order to preserve the binaural cues, two different spatial filters $\hat{\mathbf{w}}_L \in \mathbb{C}^{M \times 1}$ and $\hat{\mathbf{w}}_R \in \mathbb{C}^{M \times 1}$, are applied to the left and right hearing aid, respectively, where constraints can be used to guarantee that certain phase and magnitude relations between the left and right hearing aid outputs are preserved. Note that both spatial filters use all microphone recordings from both hearing aids.

Assume for convenience and without loss of generality that the reference microphone for the left and right hearing aid is indexed as $j = 1$ and $j = M$, respectively. In the sequel of the paper, for ease of notation, the reference terms of Eq. (1) use the subscripts $L$ and $R$ instead of $j = 1$ and $j = M$, respectively. The two enhanced output signals at the left and right hearing aids are then given by

$$\hat{x}_L = \hat{\mathbf{w}}_L^H \mathbf{y} \quad \text{and} \quad \hat{x}_R = \hat{\mathbf{w}}_R^H \mathbf{y}.$$ (4)

In Section III-A, objective measures for the preservation of binaural cues are presented. In Sections III-C—III-F, the binaural MVDR (BMVDR), the binaural LCMV (BLCMV), the optimal BLCMV (OBLCMV), and the JBLCMV are reviewed, respectively. All reviewed methods are special cases of the general binaural LCMV (GBLCMV) framework, presented in Section III-B. Finally, the basic properties of all reviewed methods are summarized in Section III-G.

### A. Binaural Cues

The extent to which the binaural cues of a specific source are preserved can be expressed using the input and output interaural tranfer function (ITF) [31]. Often the ITF is decomposed into its magnitude, describing the interaural level differences (ILDs) and its phase, describing the interaural phase differences (IPDs). The input and output ITFs of the $i$-th interferer are defined as [31]

$$\text{ITF}_{\mathbf{n}_i}^{\text{in}} = \frac{n_{iL}}{n_{iR}} = \frac{b_{iL}}{b_{iR}}, \quad \text{ITF}_{\mathbf{n}_i}^{\text{out}} = \frac{\mathbf{w}_L^H \mathbf{n}_i}{\mathbf{w}_R^H \mathbf{n}_i} = \frac{\mathbf{w}_L^H \mathbf{b}_i}{\mathbf{w}_R^H \mathbf{b}_i}.$$ (5)

The input and output ILDs are defined as [31]

$$\text{ILD}_{\mathbf{n}_i}^{\text{in}} = |\text{ITF}_{\mathbf{n}_i}^{\text{in}}|^2, \quad \text{ILD}_{\mathbf{n}_i}^{\text{out}} = |\text{ITF}_{\mathbf{n}_i}^{\text{out}}|^2.$$ (6)

The input and output IPDs are given by [31]

$$\text{IPD}_{\mathbf{n}_i}^{\text{in}} = \angle\text{ITF}_{\mathbf{n}_i}^{\text{in}}, \quad \text{IPD}_{\mathbf{n}_i}^{\text{out}} = \angle\text{ITF}_{\mathbf{n}_i}^{\text{out}}.$$ (7)

Note that frequently, the IPDs are converted and measured as time delays [32], i.e., interaural time differences (ITDs). The IPDs and ILDs are the dominant cues for binaural localization for low and high frequencies, respectively [33]. Typically, the IPDs become more important for frequencies below 1 kHz, while ILDs become more important for frequencies above 3 kHz [33]. In [34] it was experimentally shown that for broadband signals, the IPDs are perceptually much more important than the ILDs for localizing a source. More specifically, it was shown that the low frequency IPDs play the most important role perceptually for correct localization. Based on this observation several proposed multi-microphone noise reduction techniques [32], [35] leave the low frequency content of the noisy measurements unprocessed, and process only the higher frequency content. Unfortunately, if a large portion of the power of the noise is concentrated at low frequencies, the noise reduction capabilities are reduced significantly. Therefore, in this paper we aim at the simultaneous preservation of binaural cues of all sources and noise reduction at all frequencies.

A binaural spatial filter, $\mathbf{w} = [\mathbf{w}_L^T \quad \mathbf{w}_R^T]^T$, exactly preserves the binaural cues of the $i$-th interferer if $\mathrm{ITF}_{\mathbf{n}_i}^{\mathrm{in}} = \mathrm{ITF}_{\mathbf{n}_i}^{\mathrm{out}}$ [31]. Exact preservation of ITFs also implies preservation of ILDs and IPDs [31], i.e., $\mathrm{ILD}_{\mathbf{n}_i}^{\mathrm{in}} = \mathrm{ILD}_{\mathbf{n}_i}^{\mathrm{out}}$ and $\mathrm{IPD}_{\mathbf{n}_i}^{\mathrm{in}} = \mathrm{IPD}_{\mathbf{n}_i}^{\mathrm{out}}$. Non-exact preservation of binaural cues implies that there is some positive ITF error given by

$$\mathcal{E}_{\mathbf{n}_i} = |\mathrm{ITF}_{\mathbf{n}_i}^{\mathrm{out}} - \mathrm{ITF}_{\mathbf{n}_i}^{\mathrm{in}}|. \qquad (8)$$

Moreover, non-exact presevation of binaural cues implies that there is some ILD and/or IPD errors, given by

$$\mathcal{L}_{\mathbf{n}_i} = |\mathrm{ILD}_{\mathbf{n}_i}^{\mathrm{out}} - \mathrm{ILD}_{\mathbf{n}_i}^{\mathrm{in}}|, \quad \mathcal{T}_{\mathbf{n}_i} = \frac{|\mathrm{IPD}_{\mathbf{n}_i}^{\mathrm{out}} - \mathrm{IPD}_{\mathbf{n}_i}^{\mathrm{in}}|}{\pi}, \qquad (9)$$

where $0 \leq \mathcal{T}_{\mathbf{n}_i} \leq 1$ [31]. Eqs. (5), (6), (7), (8) and (9) apply also for the target source $\mathbf{x}$. As it will become obvious in the sequel, for all methods that will be discussed in this paper, the errors in Eqs. (8), (9) with respect to the target source are always zero.

As explained before, the IPD error is perceptually more important measure for binaural localization than the ILD error for broadband signals (such as speech signals contaminated by broad-band noise signals), because the IPDs are perceptually more important than the ILDs for this category of signals. Moreover, the IPD error is perceptually more informative at low frequencies, while the ILD error is perceptually more informative at high frequencies.

### B. General Binaural LCMV Framework

All binaural LCMV-based methods discussed in this section are based on a general binaural LCMV (GBLCMV)[1] framework which is the binaural version of the classical LCMV framework [7], [8]. The GBLCMV minimizes the sum of the left and right output noise powers under multiple linear equality constraints. That is,

$$\hat{\mathbf{w}}_{\mathrm{GBLCMV}} = \arg\min_{\mathbf{w} \in \mathbb{C}^{2M \times 1}} \mathbf{w}^H \tilde{\mathbf{P}} \mathbf{w} \text{ s.t. } \mathbf{w}^H \mathbf{\Lambda} = \mathbf{f}^H, \qquad (10)$$

where $\hat{\mathbf{w}}_{\mathrm{GBLCMV}} = [\hat{\mathbf{w}}_{\mathrm{GBLCMV},L}^T \quad \hat{\mathbf{w}}_{\mathrm{GBLCMV},R}^T]^T \in \mathbb{C}^{2M \times 1}$, $\mathbf{\Lambda} \in \mathbb{C}^{2M \times d}$ is assumed to be a full column rank matrix (i.e., $\mathrm{rank}(\mathbf{\Lambda}) = d$), $\mathbf{f} \in \mathbb{C}^{d \times 1}$, $d$ is the number of linear equality constraints, and

$$\tilde{\mathbf{P}} = \begin{bmatrix} \mathbf{P} & \mathbf{0} \\ \mathbf{0} & \mathbf{P} \end{bmatrix} \in \mathbb{C}^{2M \times 2M}. \qquad (11)$$

[1]We used the word *general* in order to distinguish it from the BLCMV method [26], [27].

Similarly to the classical LCMV framework [7], [8], if $d \leq 2M$, and $\mathbf{\Lambda}$ is full column rank, the GBLCMV has a closed-form solution given by

$$\hat{\mathbf{w}}_{\mathrm{GBLCMV}} = \begin{cases} \tilde{\mathbf{P}}^{-1} \mathbf{\Lambda} \left( \mathbf{\Lambda}^H \tilde{\mathbf{P}}^{-1} \mathbf{\Lambda} \right)^{-1} \mathbf{f} & \text{if } d < 2M \\ (\mathbf{\Lambda}^H)^{-1} \mathbf{f} & \text{if } d = 2M. \end{cases} \qquad (12)$$

In GBLCMV, the total number of degrees of freedom devoted to noise reduction is $\mathrm{DOF}_{\mathrm{GBLCMV}} = 2M - d$. Note that in the special case where $d = 2M$, there are no degrees of freedom left for *controlled* noise reduction, i.e., $\hat{\mathbf{w}}_{\mathrm{GBLCMV}}$ cannot reduce the objective function of the GBLCMV problem in a controlled way. Finally, if $d > 2M$, the feasible set $\{\mathbf{w} : \mathbf{w}^H \mathbf{\Lambda} = \mathbf{f}^H\}$ is empty and the GBLCMV problem has no solution. In conclusion, the matrix $\mathbf{\Lambda}$ has to be "tall" (i.e., $d < 2M$), to be able to simultaneously achieve controlled noise reduction and satisfy the constraints of the GBLCMV problem. The maximum number of constraints that the GBLCMV framework can handle, while achieving controlled noise reduction, is $d_{\max} = 2M - 1$, i.e., there should be always left at least one degree of freedom for noise reduction. Generally, the more degrees of freedom (i.e., the larger $\mathrm{DOF}_{\mathrm{GBLCMV}}$), the more controlled noise reduction can be achieved.

The set of linear constraints of the GBLCMV framework in Eq. (10) can be devided into two parts,

$$\mathbf{w}^H \left[ \mathbf{\Lambda}_1 \mid \mathbf{\Lambda}_2 \right] = \left[ \mathbf{f}_1^H \mid \mathbf{f}_2^H \right]. \qquad (13)$$

The first part consists of two distortionless constraints $\mathbf{w}_L^H \mathbf{a} = a_L$ and $\mathbf{w}_R^H \mathbf{a} = a_R$ which preserve the target source at the two reference microphones. This can be written compactly as

$$\mathbf{w}^H \mathbf{\Lambda}_1 = \mathbf{f}_1^H, \qquad (14)$$

where

$$\mathbf{\Lambda}_1 = \begin{bmatrix} \mathbf{a} & \mathbf{0} \\ \mathbf{0} & \mathbf{a} \end{bmatrix} \in \mathbb{C}^{2M \times 2}, \quad \mathbf{f}_1 = \begin{bmatrix} a_L^* \\ a_R^* \end{bmatrix} \in \mathbb{C}^{2 \times 1}.$$

All binaural methods discussed in this section are special cases of the GBLCMV framework and they share the constraints in Eq. (14), while the constraints $\mathbf{w}^H \mathbf{\Lambda}_2 = \mathbf{f}_2^H$ are different.

In the sequel of the paper we use the term $m$ ($m_{\max}$) to indicate the number (maximum number) of interferers that a special case of the GBLCMV framework can preserve, while at the same time achieving controlled noise reduction. Recall that controlled noise reduction means that there is at least one degree of freedom left for noise reduction. Moreover, $m_{\max} \leq r$ which means that some methods may be unable to preserve all simultaneously present interferers of the acoustic scene, because there are not enough available degrees of freedom.

### C. BMVDR

The BMVDR beamformer [30] can be formulated using the combination of the following two beamformers

$$\hat{\mathbf{w}}_{\mathrm{BMVDR},L} = \arg\min_{\mathbf{w}_L \in \mathbb{C}^{M \times 1}} \mathbf{w}_L^H \mathbf{P} \mathbf{w}_L \text{ s.t. } \mathbf{w}_L^H \mathbf{a} = a_L, \qquad (15)$$

$$\hat{\mathbf{w}}_{\mathrm{BMVDR},R} = \arg\min_{\mathbf{w}_R \in \mathbb{C}^{M \times 1}} \mathbf{w}_R^H \mathbf{P} \mathbf{w}_R \text{ s.t. } \mathbf{w}_R^H \mathbf{a} = a_R, \qquad (16)$$

with closed-form solutions

$$\hat{\mathbf{w}}_{\text{BMVDR},L} = \frac{\mathbf{P}^{-1}\mathbf{a}a_L^*}{\mathbf{a}^H\mathbf{P}^{-1}\mathbf{a}}, \quad \hat{\mathbf{w}}_{\text{BMVDR},R} = \frac{\mathbf{P}^{-1}\mathbf{a}a_R^*}{\mathbf{a}^H\mathbf{P}^{-1}\mathbf{a}}. \quad (17)$$

The BMVDR is the simplest special case of the GBLCMV framework in the sense that it has the minimum number of constraints ($d = 2$) given by Eq. (14). Specifically, the two optimization problems in Eqs. (15) and (16) can be reformulated as the following joint optimization problem,

$$\hat{\mathbf{w}}_{\text{BMVDR}} = \underset{\mathbf{w}\in\mathbb{C}^{2M\times 1}}{\arg\min} \mathbf{w}^H\tilde{\mathbf{P}}\mathbf{w} \text{ s.t. } \mathbf{w}^H\mathbf{\Lambda}_1 = \mathbf{f}_1^H, \quad (18)$$

where $\hat{\mathbf{w}}_{\text{BMVDR}} = [\hat{\mathbf{w}}_{\text{BMVDR},L}^T \quad \hat{\mathbf{w}}_{\text{BMVDR},R}^T]^T \in \mathbb{C}^{2M\times 1}$. Since, the BMVDR algorithm has the minimum possible number of constraints, the total number of degrees of freedom which can be devoted to noise reduction is $\text{DOF}_{\text{BMVDR}} = 2M - 2$.

The BMVDR beamformer preserves the binaural cues of the target source, but distorts the binaural cues of all the interferers [30], i.e., $m_{\max} = 0$. More specifically, after beamforming, the binaural cues of the interferers collapse on the binaural cues of the target source. It can be easily shown [30] that the binaural cues of the target source are preserved due to the satisfaction of the two distortionless constraints of the problems in Eqs. (15) and (16). That is,

$$\text{ITF}_{\mathbf{x}}^{\text{in}} = \text{ITF}_{\mathbf{x}}^{\text{out}} = \frac{a_L}{a_R}, \quad (19)$$

Therefore, the ITF error is $\mathcal{E}_{\mathbf{x},\text{BMVDR}} = 0$. Furthermore, it can be easily shown that the binaural cues of the interferers collapse to the binaural cues of the target source [30]. More specifically, the $\text{ITF}_{\mathbf{n}_i}^{\text{in}}$ is given by

$$\text{ITF}_{\mathbf{n}_i}^{\text{in}} = \frac{b_{iL}}{b_{iR}}, \quad (20)$$

while $\text{ITF}_{\mathbf{n}_i}^{\text{out}}$ is given by

$$\text{ITF}_{\mathbf{n}_i}^{\text{out}} = \frac{\hat{\mathbf{w}}_{\text{BMVDR},L}^H\mathbf{b}_i}{\hat{\mathbf{w}}_{\text{BMVDR},R}^H\mathbf{b}_i} = \frac{\frac{\mathbf{a}^H\mathbf{P}^{-1}\mathbf{b}_ia_L}{\mathbf{a}^H\mathbf{P}^{-1}\mathbf{a}}}{\frac{\mathbf{a}^H\mathbf{P}^{-1}\mathbf{b}_ia_R}{\mathbf{a}^H\mathbf{P}^{-1}\mathbf{a}}} = \frac{a_L}{a_R} = \text{ITF}_{\mathbf{x}}^{\text{in}}. \quad (21)$$

Thus, after beamforming, the interferers will have the same ITF as the target source and their ITF error is given by

$$\mathcal{E}_{\mathbf{n}_i,\text{BMVDR}} = \left|\text{ITF}_{\mathbf{n}_i}^{\text{out}} - \text{ITF}_{\mathbf{n}_i}^{\text{in}}\right| = \left|\frac{a_L}{a_R} - \frac{b_{iL}}{b_{iR}}\right|. \quad (22)$$

### D. BLCMV

Another special case of the GBLCMV framework is the binaural linearly constrained minimum variance (BLCMV) beamformer [26], [27] which, unlike the BMVDR, uses additional constraints for the preservation of the binaural cues of $m$ interferers. The left and right spatial filters of the BLCMV are given by [26], [27]

$$\hat{\mathbf{w}}_{\text{BLCMV},L} = \underset{\mathbf{w}_L\in\mathbb{C}^{M\times 1}}{\arg\min} \mathbf{w}_L^H\mathbf{P}\mathbf{w}_L$$
$$\text{s.t.} \quad \mathbf{w}_L^H\mathbf{a} = a_L$$
$$\mathbf{w}_L^H\mathbf{b}_1 = \eta_Lb_{1L}, \ \ldots, \ \mathbf{w}_L^H\mathbf{b}_m = \eta_Lb_{mL}, \quad (23)$$

and

$$\hat{\mathbf{w}}_{\text{BLCMV},R} = \underset{\mathbf{w}_R\in\mathbb{C}^{M\times 1}}{\arg\min} \mathbf{w}_R^H\mathbf{P}\mathbf{w}_R$$
$$\text{s.t.} \quad \mathbf{w}_R^H\mathbf{a} = a_R$$
$$\mathbf{w}_R^H\mathbf{b}_1 = \eta_Lb_{1R}, \ \ldots, \ \mathbf{w}_R^H\mathbf{b}_m = \eta_Lb_{mR}, \quad (24)$$

where the constraints $\mathbf{w}_L^H\mathbf{a} = a_L$ and $\mathbf{w}_R^H\mathbf{a} = a_R$ are the two common distortionless constraints used in all special cases in the GBLCMV framework, while the constraints $\mathbf{w}_L^H\mathbf{b}_i = \eta_Lb_{iL}$ and $\mathbf{w}_R^H\mathbf{b}_i = \eta_Rb_{iR}$, for $i = 1, \ldots, m$, aim at a) preserving the binaural cues and b) supressing the $m$ interferers. The amount of supression is controlled via the interference rejection parameters $\eta_L$ and $\eta_R$ which are pre-defined ($0 \leq \eta_L$, $\eta_R < 1$) real-valued scalars. Binaural cue preservation is achieved only if $\eta = \eta_L = \eta_R$ [26], [28]. The two optimization problems in Eqs. (23) and (24) can be compactly formulated as a joint optimization problem. That is,

$$\hat{\mathbf{w}}_{\text{BLCMV}} = \underset{\mathbf{w}\in\mathbb{C}^{2M\times 1}}{\arg\min} \mathbf{w}^H\tilde{\mathbf{P}}\mathbf{w} \text{ s.t. } \mathbf{w}^H\mathbf{\Lambda} = \mathbf{f}^H, \quad (25)$$

where

$$\mathbf{\Lambda} = \begin{bmatrix} \mathbf{\Lambda}_1 & \mathbf{\Lambda}_2 \end{bmatrix} = \underbrace{\begin{bmatrix} \mathbf{a} & \mathbf{0} & \mathbf{b}_1 & \mathbf{0} & \cdots & \mathbf{b}_m & \mathbf{0} \\ \mathbf{0} & \mathbf{a} & \mathbf{0} & \mathbf{b}_1 & \cdots & \mathbf{0} & \mathbf{b}_m \end{bmatrix}}_{\mathbb{C}^{2M\times(d=2+2m)}},$$

and

$$\mathbf{f}^T = \begin{bmatrix} \mathbf{f}_1^T & \mathbf{f}_2^T \end{bmatrix}$$
$$= \underbrace{\begin{bmatrix} a_L^* & a_R^* & \eta_Lb_{1L}^* & \eta_Rb_{1R}^* & \cdots & \eta_Lb_{mL}^* & \eta_Rb_{mR}^* \end{bmatrix}}_{\mathbb{C}^{1\times(d=2+2m)}}.$$

The available degrees of freedom for noise reduction are $\text{DOF}_{\text{BLCMV}} = 2M - d = 2M - 2m - 2$. Since $d_{\max} = 2M - 1$ (see Section III-B), BLCMV can simultaneously achieve controlled noise suppression and binaural cue preservation of at most $m_{\max} = M - 2$ interferers.

The ITF errors of the target source and of the $m$ interferers that are included in the constraints are zero, i.e., $\mathcal{E}_{\mathbf{x},\text{BLCMV}} = 0$ and $\mathcal{E}_{\mathbf{n}_i,\text{BLCMV}} = 0$, for $i = 1, \ldots, m \leq r$. However, if some interferers are not included in the constraints, their ITF error will be non-zero, i.e., $\mathcal{E}_{\mathbf{n}_i,\text{BLCMV}} > 0$, for $i = m + 1, \cdots, r$.

### E. OBLCMV

The OBLCMV [28] can be seen as a special case of the BLCMV (and, hence, the GBLCMV) since it solves the same optimization problem. However, it preserves the binaural cues of only one interferer (e.g., the $k$-th interferer) using an optimal complex-valued interference rejection parameter $\hat{\eta} = \hat{\eta}_L = \hat{\eta}_R$ with respect to the binaural output SNR (defined in Sec. VI-B2). More specifically, OBLCMV solves the optimization problem in Eq. (25) where $\mathbf{\Lambda}$ and $\mathbf{f}^T$, are given by [28]

$$\mathbf{\Lambda} = \begin{bmatrix} \mathbf{\Lambda}_1 & \mathbf{\Lambda}_2 \end{bmatrix} = \begin{bmatrix} \mathbf{a} & \mathbf{0} & \mathbf{b}_1 & \mathbf{0} \\ \mathbf{0} & \mathbf{a} & \mathbf{0} & \mathbf{b}_1 \end{bmatrix} \in \mathbb{C}^{2M\times 4},$$

$$\mathbf{f}^T = \begin{bmatrix} \mathbf{f}_1^T & \mathbf{f}_2^T \end{bmatrix} = \begin{bmatrix} a_L^* & a_R^* & \hat{\eta}b_{kL}^* & \hat{\eta}b_{kR}^* \end{bmatrix} \in \mathbb{C}^{1\times 4} \quad (26)$$

where $1 \le k \le r$. The available degrees of freedom for noise reduction are $\text{DOF}_{\text{OBLCMV}} = 2M - 4$.

The ITF errors of the target source and of the $k$-th interferer that are included in the constraints are zero, i.e., $\mathcal{E}_{\mathbf{x}, \text{OBLCMV}} = 0$ and $\mathcal{E}_{\mathbf{n}_k, \text{OBLCMV}} = 0$. However, the binaural cues of all the other $r - 1$ interferers will be distorted, i.e., $\mathcal{E}_{\mathbf{n}_i, \text{BLCMV}} > 0$, for $i \in \{1, \cdots, r\} - \{k\}$.

### F. JBLCMV

Recall from Section III-A that preserving binaural cues of the $i$-th interferer implies that the following constraint has to be satisfied

$$\text{ITF}_{\mathbf{n}_i}^{\text{in}} = \text{ITF}_{\mathbf{n}_i}^{\text{out}} \implies \frac{\mathbf{w}_L^H \mathbf{b}_i}{\mathbf{w}_R^H \mathbf{b}_i} = \frac{b_{iL}}{b_{iR}}, \quad (27)$$

which can be reformulated as:

$$\mathbf{w}_L^H \mathbf{b}_i b_{iR} - \mathbf{w}_R^H \mathbf{b}_i b_{iL} = 0. \quad (28)$$

Compared to (O)BLCMV this unified constraint reduces the number of constraints, used for binaural cue preservation, by a factor 2. As a result, for a given number of interferers, more degrees of freedom can be devoted to noise reduction. The JBLCMV [29], [30] uses this type of equality constraints for the preservation of the binaural cues of $m$ interferers. More specifically, the JBLCMV problem is given by

$$\hat{\mathbf{w}}_{\text{JBLCMV}} = \underset{\mathbf{w} \in \mathbb{C}^{2M \times 1}}{\arg \min} \ \mathbf{w}^H \tilde{\mathbf{P}} \mathbf{w} \ \text{s.t.} \ \mathbf{w}^H \boldsymbol{\Lambda} = \mathbf{f}^H, \quad (29)$$

where

$$\begin{aligned} \boldsymbol{\Lambda} &= [\boldsymbol{\Lambda}_1 \mid \boldsymbol{\Lambda}_2] \\ &= \begin{bmatrix} \mathbf{a} & \mathbf{0} & \mathbf{b}_1 b_{1R} & \cdots & \mathbf{b}_m b_{mR} \\ \mathbf{0} & \mathbf{a} & -\mathbf{b}_1 b_{1L} & \cdots & -\mathbf{b}_m b_{mL} \end{bmatrix} \in \mathbb{C}^{2M \times (2+m)} \end{aligned} \quad (30)$$

and $\mathbf{w}_{\text{JBLCMV}} = [\mathbf{w}_{\text{JBLCMV},L}^T \quad \mathbf{w}_{\text{JBLCMV},R}^T]^T$. Moreover,

$$\begin{aligned} \mathbf{f}^T &= [\mathbf{f}_1^T \mid \mathbf{f}_2^T] \\ &= [a_L^* \quad a_R^* \mid 0 \quad 0 \quad \cdots \quad 0] \in \mathbb{C}^{1 \times (2+m)}. \end{aligned} \quad (31)$$

Similarly to all other special cases of the GBLCMV framework, $\mathbf{w}^H \boldsymbol{\Lambda}_1 = \mathbf{f}_1^H$ is used for the exact binaural cue preservation of the target source, while $\mathbf{w}^H \boldsymbol{\Lambda}_2 = \mathbf{f}_2^H$ is used for the preservation of the binaural cues of $m$ interferers.

The JBLCMV can simultaneously achieve controlled noise reduction and binaural cue preservation of up to $m_{\max} = 2M - 3$ interferers [29]. Moreover, the degrees of freedom devoted to noise reduction is $\text{DOF}_{\text{JBLCMV}} = 2M - m - 2$.

### G. Summary of GBLCMV methods

We summarize some of the properties of the methods discussed in Section III. Table I gives an overview of two important factors: a) the maximum number of interferers' binaural cues that can be preserved while achieving controlled noise reduction $m_{\max}$, and b) the degrees of freedom (DOF) available for noise reduction. The following conclusions can be drawn from this table:

- The BMVDR has the maximum DOF, which means that it can achieve the best possible noise reduction. It

| Method | $m_{\max}$ | DOF |
|---|---|---|
| BMVDR [30] | 0 | $2M - 2$ |
| BLCMV [27] | $M - 2$ | $2M - 2m - 2$ |
| OBLCMV [28] | 1 | $2M - 4$ |
| JBLCMV [29], [30] | $2M - 3$ | $2M - m - 2$ |

preserves the binaural cues of the target source, but not the binaural cues of the interferers.
- Unlike (O)BLCMV which uses two constraints per interferer, JBLCMV uses only one constraint per interferer. Therefore, JBLCMV can preserve the binaural cues of more interferers, or equivalently, given the same number of interferers it has more available degrees of freedom devoted to noise reduction.

In this paper, if the number of simultaneously present interferers is $r > m_{\max}$, the extra interferers $r - m_{\max}$ are *not* included in the constraints in the GBLCMV methods, in order to always have one degree of freedom left for controlled noise reduction.

## IV. PROPOSED NON-CONVEX PROBLEM

In this section, we present a general optimization problem of which BMVDR and JBLCMV are special cases. More specifically, we relax the constraints on the binaural cues of the interferers, while keeping the strict equality constraints on the target source (i.e., $\mathbf{w}^H \boldsymbol{\Lambda}_1 = \mathbf{f}_1^H$). The relaxation allows to trade-off the amount of noise reduction and binaural cue preservation per interferer in a controlled way. The proposed optimization problem is defined as

$$\begin{aligned} \hat{\mathbf{w}} = \underset{\mathbf{w} \in \mathbb{C}^{2M \times 1}}{\arg \min} \ & \mathbf{w}^H \tilde{\mathbf{P}} \mathbf{w} \ \text{s.t.} \ \mathbf{w}^H \boldsymbol{\Lambda}_1 = \mathbf{f}_1^H, \\ & \underbrace{\left| \frac{\mathbf{w}_L^H \mathbf{b}_i}{\mathbf{w}_R^H \mathbf{b}_i} - \frac{b_{iL}}{b_{iR}} \right|}_{\mathcal{E}_{\mathbf{n}_i}} \le e_i, \quad i = 1, \cdots, m. \end{aligned} \quad (32)$$

The inequality constraints bound the ITF error (see Eq. (8)), for the interferers $i = 1, \cdots, m$ to be less than a positive trade-off parameter $e_i, i = 1, \cdots, m$. These inequality constraints will be transformed, in the sequel of this section (see Eqs. (34), (35)), in such a way that they can be viewed as relaxations of the strict equality constraints in Eq. (28) used in the JBLCMV method. Note that the proposed method is flexible to choose a different $e_i$ for every interferer according to its importance. For instance, maybe certain locations are more important to be preserved than others and, therefore, a smaller $e_i$ must be used. The trade-off parameter $e_i$, is selected as

$$e_i(c_i) = c_i \mathcal{E}_{\mathbf{n}_i, \text{BMVDR}}, \quad (33)$$

where $0 \leq c_i \leq 1$ controls the amount of binaural cue collapse towards the target source, and the amount of noise reduction of the $i$-th interferer. If $c_i = 1, \forall i$ is used in the optimization problem in Eq. (32), then $\hat{\mathbf{w}} = \hat{\mathbf{w}}_{\text{BMVDR}}$ which is seen as a worst case, with respect to binaural cue preservation, because there is total collapse of binaural cues of the interferers towards the binaural cues of the target source. If $c_i = 0, \forall i$ we have perfect preservation of binaural cues of the $m$ interferers, and $\hat{\mathbf{w}} = \hat{\mathbf{w}}_{\text{JBLCMV}}$. Without any loss of generality, for notational convenience, we assume that the binaural cues of all interferers are of equal importance and, therefore, $c_i = c, \forall i$. Moreover, we keep $c$ fixed over all frequency bins. It is worth noting that other strategies for choosing $c$ may exist, which might lead to a better trade-off between maximum possible noise reduction and perceptual binaural cue preservation. As explained in Section III-A, low frequency content is perceptually more important for binaural cue preservation than high frequency content. Thus, smaller $c$ values for low frequencies and larger $c$ values for higher frequencies may give a better perceptual trade-off.

The problem in Eq. (32) is not a convex problem and it is hard to solve. In Section V we propose a method that approximately solves the non-convex problem in an iterative way by solving at each iteration a convex problem.

## V. Proposed Iterative Convex Problem

By doing some simple algebraic manipulations, the optimization problem in Eq. (32) can equivalently be written as

$$\hat{\mathbf{w}} = \arg\min_{\mathbf{w} \in \mathbb{C}^{2M \times 1}} \mathbf{w}^H \tilde{\mathbf{P}} \mathbf{w} \text{ s.t. } \mathbf{w}^H \mathbf{\Lambda}_1 = \mathbf{f}_1^H,$$
$$\frac{|\mathbf{w}_L^H \mathbf{b}_i b_{iR} - \mathbf{w}_R^H \mathbf{b}_i b_{iL}|}{|\mathbf{w}_R^H \mathbf{b}_i b_{iR}|} \leq e_i(c),$$
$$\text{for } i = 1, \cdots, m. \quad (34)$$

Furthermore, the optimization problem in Eq. (34) can be re-written as

$$\hat{\mathbf{w}} = \arg\min_{\mathbf{w} \in \mathbb{C}^{2M \times 1}} \mathbf{w}^H \tilde{\mathbf{P}} \mathbf{w} \text{ s.t. } \mathbf{w}^H \mathbf{\Lambda}_1 = \mathbf{f}_1^H,$$
$$|\mathbf{w}^H \mathbf{\Lambda}_{2,i}| \leq \underbrace{|e_i(c)\mathbf{w}_R^H \mathbf{b}_i b_{iR}|}_{f_{2,i}},$$
$$\text{for } i = 1, \cdots, m, \quad (35)$$

where $\mathbf{\Lambda}_{2,i}$ is the $i$-th column of $\mathbf{\Lambda}_2$ in Eq. (30).

We approximately solve the non-convex problem in Eq. (35) in an iterative way using $\mathbf{w}_R^H$ of the previous iteration in $f_{2,i}, i = 1, \cdots, m$. The new iterative problem is convex at each iteration and is given by

$$\hat{\mathbf{w}}_{(k)} = \arg\min_{\mathbf{w} \in \mathbb{C}^{2M \times 1}} \mathbf{w}^H \tilde{\mathbf{P}} \mathbf{w} \text{ s.t. } \mathbf{w}^H \mathbf{\Lambda}_1 = \mathbf{f}_1^H,$$
$$|\mathbf{w}^H \mathbf{\Lambda}_{2,i}| \leq \underbrace{|e_i(c)\hat{\mathbf{w}}_{R,(k-1)}^H \mathbf{b}_i b_{iR}|}_{f_{2,i,(k)}}$$
$$\text{for } i = 1, \cdots, m, \quad (36)$$

where $\hat{\mathbf{w}}_{(k)} = [\hat{\mathbf{w}}_{L,(k)}^T \quad \hat{\mathbf{w}}_{R,(k)}^T]^T$ is the estimated binaural spatial filter of the $k$-th iteration, which is initialized as $\hat{\mathbf{w}}_{(0)} = \hat{\mathbf{w}}_{\text{BMVDR}}$.

Similarly to other existing minimum variance beamformers with inequality constraints [36], [37], the convex optimization problem in Eq. (36) can be equivalently written as a second order cone programming (SOCP) problem with equality and inequality constraints (see Appendix) and it can be solved efficiently with interior point methods [38].

The ITF error of the $i$-th interferer at the $k$-th iteration is given by

$$\mathcal{E}_{\mathbf{n}_i,(k)} = \left| \frac{\hat{\mathbf{w}}_{L,(k)}^H \mathbf{b}_i}{\hat{\mathbf{w}}_{R,(k)}^H \mathbf{b}_i} - \frac{b_{iL}}{b_{iR}} \right|. \quad (37)$$

This iterative method is stopped when all the constraints of the original problem in Eq. (32) are satisfied. The stopping criterion that we use for the proposed iterative method is given by

$$\mathcal{E}_{\mathbf{n}_i,(k)} \leq e_i(c), \text{ for } i = 1, \cdots, m, \quad (38)$$

where $e_i(c)$ is given in Eq. (33). Recall that $\mathbf{f}_2 = \mathbf{0}$ (i.e., $f_{2,i} = 0, \forall i$) is used in JBLCMV. Unlike JBLCMV, the proposed method uses $f_{2,i,(k)} \geq 0, \forall i$ and, therefore, the constraints dedicated for the preservation of binaural cues of the interferers are seen as relaxations of the strict equality constraints of the JBLCMV method. These relaxations enlarge the feasible set of the problem, allowing more constraints to be used compared to JBLCMV. The JBLCMV can be seen as a special case of the proposed method for $c = 0$, $f_{2,i,(1)} = 0, i = 1, \cdots m$. In this case, the relaxed constraints in the proposed method become identical to the strict constraints of the JBLCMV. Hence, the JBLCMV needs to run only one iteration of the problem in Eq. (36). If $c = 0$, the proposed method follows the same strategy for handling $r > m_{\max}$ simultaneously present interferers as in Section III-G. However, if $c > 0$, then there is a typically large, difficult to predict $m_{\max}$[2], due to the inequality constraints and, therefore, the proposed method uses $m = r, \forall r$ constraints for the preservation of the binaural cues of all simultaneously present interferers. Finally, if $c = 1$, the proposed method does not iterate and stops immediately giving as output the initialization $\hat{\mathbf{w}}_{(0)} = \hat{\mathbf{w}}_{\text{BMVDR}}$.

The termination of the proposed iterative method *may* need a large amount of iterations because of the fixed $c$ in Eq. (36). The reason for this is explained in detail in Section V-A. To control the speed of termination we replace in Section V-B the fixed $c$ in Eq. (36) with a decreasing parameter $\tau_{(k)}$ (initialized with $\tau_{(0)} = c$) which controls the speed of termination. In Section V-C we show under which conditions the proposed algorithm guarantees that it will find a feasible solution satisfying the stopping criterion in Eq. (38) in a finite number of iterations. An overview of the proposed method using the adaptive $\tau_{(k)}$ is given in Algorithm 1.

### A. Speed of Termination

The proposed iterative method may have slow termination due to the fixed choice of $c$. In this section we explain the reason and in Section V-B we explain how to control the speed of termination.

---

[2]The feasible set of the proposed method typically reduces by adding more inequality constraints. However it is difficult to predict after how many constraints, $m$, it becomes empty, i.e., what is the value of $m_{\max}$.

Let $\Phi_{(k)}$ denote the convex feasible set in the $k$-th iteration of the iterative optimization problem in Eq. (36) given by

$$\Phi_{(k)} = \bigcap_{i=1}^{m} \left\{ \mathbf{w}_{(k)} : \boldsymbol{\Lambda}_1^H \mathbf{w}_{(k)} = \mathbf{f}_1, |\mathbf{w}_{(k)}^H \boldsymbol{\Lambda}_{2,i}| \le f_{2,i,(k)} \right\}, \quad (39)$$

and $\Psi(c)$ the non-convex feasible set of the original non-convex problem of Eqs. (32), (33) given by

$$\Psi(c) = \bigcap_{i=1}^{m} \left\{ \mathbf{w} : \boldsymbol{\Lambda}_1^H \mathbf{w} = \mathbf{f}_1, \left| \frac{\mathbf{w}_L^H \mathbf{b}_i}{\mathbf{w}_R^H \mathbf{b}_i} - \frac{b_{iL}}{b_{iR}} \right| \le c \mathcal{E}_{\mathbf{n}_i, \text{BMVDR}} \right\}, \quad (40)$$

where $\hat{\mathbf{w}}_{\text{JBLCMV}} \in \Psi(0)$, and $\Psi(0) \subseteq \Psi(c), 0 \le c \le 1$ and, therefore, $\hat{\mathbf{w}}_{\text{JBLCMV}} \in \Psi(c), 0 \le c \le 1$. In words, $\hat{\mathbf{w}}_{\text{JBLCMV}}$ is the feasible solution that belongs to $\Psi(0)$ and gives the minimum output noise power.

Note that the $\Phi_{(k)}$ changes for every next iteration, while $\Psi(c)$ is constant over time. We can think of $\Phi_{(k)}$ as a convex approximation set of $\Psi(c)$ at iteration $k$ (see a simplistic example of the two sets in Fig. 1(a)).

Note that the proposed iterative method will typically try to find a solution on the boundary of $\Phi_{(k)}$. Some parts of the boundary of $\Phi_{(k)}$ will be inside or on the boundary of $\Psi(c)$, while other parts can be outside the set $\Psi(c)$. Therefore, it is possible that the estimated $\hat{\mathbf{w}}_{(k)}$ will be outside of $\Psi(c)$ (see Fig. 1(a) for instance). In this case, obviously, the stopping criterion is not satisfied and, therefore, the problem goes to the next iteration. In the next iteration, $\Phi_{(k+1)}$ changes and a new $\hat{\mathbf{w}}_{(k+1)}$ is estimated which can be again outside of $\Psi(c)$ (see Fig. 1(a) for instance). This repetition can happen many times leading to a very slow termination because the new estimate $\hat{\mathbf{w}}_{(k+1)}$ is not selected according to a binaural-cue error descent direction. To avoid this undesirable situation, we propose in Section V-B to replace the fixed $c$ in Eq. (36) with an adaptive reduction parameter $\tau_{(k)}$, in order to make sure that solutions that are on the boundary of $\Phi_{(k)}$ and that are outside $\Psi(c)$ will progressively provide a reduced binaural-cue error, i.e., to move towards the direction of the interior of $\Psi(c)$ (see Fig. 1(b) for instance).

### B. Avoiding Slow Termination

The termination of the proposed iterative method *may* need a large amount of iterations because of the fixed $c$ in Eq. (36), as explained in Section V-A. Therefore, the replacement of $c$ with an adaptive reduction parameter $\tau_{(k)}$ *only* in Eq. (36) is useful for guaranteed termination within a *pre-selected* finite maximum number of iterations, $k_{\max}$. More specifically, the new adaptive reduction parameter that we use in Eq. (36) instead of $c$ is given by

$$\tau_{(k)} = \tau_{(k-1)} - \alpha_{(k_{\max})}, \quad (41)$$

where $\tau_{(0)} = c$ is selected according to the initial desired amount of collapse of binaural cues in the original non-convex problem in Eqs. (32), (33). The step $\alpha_{(k_{\max})}$ controls the speed of termination (i.e., how fast the stopping criterion will be satisfied), and is a function of the maximum allowed number of iterations for termination given from the user. That is,

$$\alpha_{(k_{\max})} = \frac{c}{k_{\max}}. \quad (42)$$

---

**Algorithm 1** Proposed Iterative Method

**Input:** $c, k_{\max}, \mathbf{a}, \mathbf{b}_i, i = 1, \cdots, m$
**Output:** $\hat{\mathbf{w}}_{(k)}$
    *Initialisation* : $\hat{\mathbf{w}}_{(0)} \leftarrow \hat{\mathbf{w}}_{\text{BMVDR}}, k \leftarrow 1, \tau_{(0)} \leftarrow c$
    *General comments* :
        {SC stands for stopping criterion in Eq. (38)}.
        {SP stands for solving problem in Eq. (36)}.
1: **if** SC($\hat{\mathbf{w}}_{(0)}, c$) = **true then**
2:     go to 17
3: **end if**
    *start iterations*
4: **while** $k \le k_{\max}$ **do**
5:     **if** $k = k_{\max}$ **then**
6:         $\hat{\mathbf{w}}_{(k)} \leftarrow \text{SP} \left( \hat{\mathbf{w}}_{(k-1)}, \tau_{(k)}, \mathbf{a}, \mathbf{b}_i, i = 1, \cdots, 2M - 3 \right)$
7:         go to 17
8:     **else**
9:         $\hat{\mathbf{w}}_{(k)} \leftarrow \text{SP} \left( \hat{\mathbf{w}}_{(k-1)}, \tau_{(k)}, \mathbf{a}, \mathbf{b}_i, i = 1, \cdots, m \right)$
10:    **end if**
11:    **if** SC($\hat{\mathbf{w}}_{(k)}, c$) = **true then**
12:        go to 17
13:    **end if**
14:    $k \leftarrow k + 1$
15:    $\tau_{(k)} = \tau_{(k-1)} - c/k_{\max}$
16: **end while**
17: **return** $\hat{\mathbf{w}}_{(k)}$

---

Note that we replace $c$ with $\tau_{(k)}$ only in Eq. (36) and not in the stopping criterion in Eq. (38). This is because, the stopping criterion is based on the fixed feasible set $\Psi(c)$ of the non-convex problem in Eq. (32) which should remain constant over iterations (see an example of two consecutive iterations in Fig. 1). Moreover, the $\tau_{(k)}$ is always non-negative, because $\tau_{(k_{\max})} = 0$. Small $k_{\max}$, speeds up the reduction of $\tau_{(k)}$ and, thus, it also speeds up the termination of the proposed method. Of course a very small $k_{\max}$ can lead to a feasible solution, $\hat{\mathbf{w}}_{(k)}$, for which $\sum_i \mathcal{E}_{\mathbf{n}_i,(k)} \ll \sum_i e_i(c)$, i.e., to be far away from the boundary of $\Psi(c)$. This means that $\hat{\mathbf{w}}_{(k)}$ provides better binaural cue preservation than the desired amount of binaural cue preservation, $e_i(c)$. As a result, there will be less noise suppression. Ideally, we would like to arrive as close as possible to the controlled trade-off between noise reduction and binaural cue preservation given by our initial specifications (i.e., amount of collapse). Therefore, a careful choice of $k_{\max}$ is needed in order to find a feasible solution $\hat{\mathbf{w}}_{(k)}$ that:

- achieves a binaural-cue error $\sum_i \mathcal{E}_{\mathbf{n}_i,(k)} \approx \sum_i e_i(c)$, i.e., to be as close as possible to the boundary of $\Psi(c)$.
- to terminate as fast as possible.

Of course there is a trade-off between the two goals.

### C. Guaranteed Termination

In this section, we prove that the proposed iterative method using the adaptive reduction parameter in Eq. (41) *guarantees* termination, simultaneous controlled approximate binaural cue preservation, and controlled noise reduction, in *at most* $k_{\max}$ iterations, for a limited number of interferers $m \le 2M - 3$.

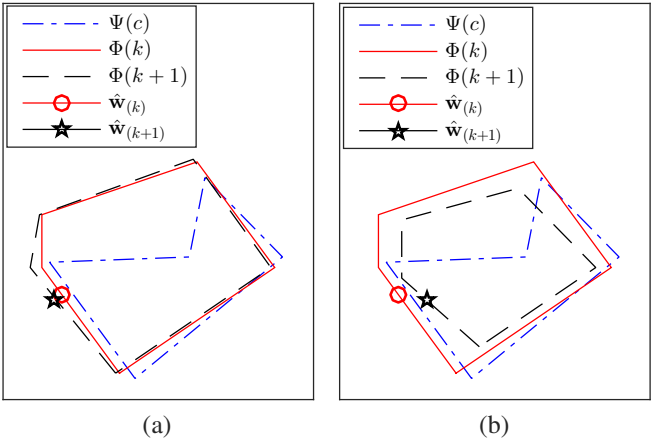

(a)                    (b)

Fig. 1. Simplistic visualization of two successive iterations ($k$ and $k + 1$) of the proposed method with (a) a fixed $c$, (b) a reducing $\tau(k)$. In $k+1$ iteration the stopping criterion is satisfied in (b). On the contrary, in (a) the stopping criterion is not satisfied, because $\hat{\mathbf{w}}_{(k+1)} \notin \Psi(c)$.

Nevertheless, our simulation experiments (see Section VI-C) show that our algorithm a) is capable of simultaneously achieving controlled approximate binaural cue preservation and, in most cases, controlled noise reduction of *more* interferers than $2M - 3$ for $c > 0$, and b) finds a feasible solution in much fewer iterations, on average, than $k_{max}$, for $k_{max} = 10, 50$.

The adaptive decreasing of $\tau_{(k)}$ (see Eq. (41)) results in an adaptive shrinking of $\Phi_{(k)}$. Therefore, in the case where the estimated $\hat{\mathbf{w}}_{(k)}$ will be outside of $\Psi(c)$, the stopping criterion is not satisfied and, therefore, the algorithm continues with the next iteration. In the next iteration, $\Phi_{(k)}$ typically shrinks due to the decreased value of $\tau_{(k)}$ according to Eq. (41). The algorithm continues until there is a solution $\hat{\mathbf{w}}_{(k)} \in \Psi(c)$. Note that this does not necessarily mean that the algorithm will stop if and only if $\Phi_{(k)} \subseteq \Psi(c)$ (see e.g., Fig. 1(b) where the algorithm stops before $\Phi_{(k)} \subseteq \Psi(c)$). Only in the worst case scenario a solution is found when $\Phi_{(k)} \subseteq \Psi(c)$. We show below that, for $m \leq 2M - 3$, the proposed method guarantees termination within a pre-defined finite maximum number of iterations, $k_{max}$, while achieving controlled binaural cue preservation accuracy and controlled noise reduction. This is written more formally in the following theorem.

*Theorem 1:* If $m \leq 2M - 3$, the proposed method a) will always find a solution in a finite number of iterations $k \leq k_{max}$ satisfying the stopping criterion of Eq. (38), and b) will always have a bounded ITF error, i.e.,

$$0 \leq \mathcal{E}_{\mathbf{n}_i,(k)} \leq e_i(c), \text{ for } i = 1, \cdots, m, \quad (43)$$

and a bounded noise output power

$$\hat{\mathbf{w}}_{BMVDR}^H \tilde{\mathbf{P}} \hat{\mathbf{w}}_{BMVDR} \leq \hat{\mathbf{w}}_{(k)}^H \tilde{\mathbf{P}} \hat{\mathbf{w}}_{(k)} \leq \hat{\mathbf{w}}_{JBLCMV}^H \tilde{\mathbf{P}} \hat{\mathbf{w}}_{JBLCMV}. \quad (44)$$

Proof: Note that for $m \leq 2M - 3$, after $k_{max}$ iterations $\tau_{(k)} = 0$ (see Eq. 41) and, therefore, $\hat{\mathbf{w}}_{(k_{max})} = \hat{\mathbf{w}}_{JBLCMV}$ because the relaxations of the proposed method in Eq. (36) become $\mathbf{w}_{(k)}^H \mathbf{\Lambda}_2 = 0$, which is the same as in JBLCMV as explained in Section V. Therefore, for $m \leq 2M - 3$, the algorithm, in the worst case scenario, will terminate after $k_{max}$ iterations (specified by the user), giving the solution $\hat{\mathbf{w}}_{JBLCMV}$ which always satisfies the stopping criterion, i.e., $\hat{\mathbf{w}}_{JBLCMV} \in \Psi(c)$, for $0 \leq c \leq 1$ (see Section V-A). This means that the algorithm in the worst case scenario (after $k_{max}$) will have the noise output power $\hat{\mathbf{w}}_{JBLCMV}^H \tilde{\mathbf{P}} \hat{\mathbf{w}}_{JBLCMV}$ and an ITF error equal 0. Moreover, the noise output power cannot be less than $\hat{\mathbf{w}}_{BMVDR}^H \tilde{\mathbf{P}} \hat{\mathbf{w}}_{BMVDR}$ (because $\hat{\mathbf{w}}_{BMVDR}$ achieves the best noise reduction over all the aforementioned methods, because it has the largest feasible set) and the ITF error will be $\mathcal{E}_{\mathbf{n}_i,(k)} \leq e_i(c) \forall i$, because the stopping criterion is satisfied. Therefore, Eqs. (43) and (44) are proved.

Note that, for $k = k_{max}$ and $m > 2M - 2$, $\Phi_{(k_{max})} = \emptyset$[3]. However, for $k < k_{max}$ and $m > 2M - 2$, $\Phi_{(k)}$ may not be empty. As we will show in our experiments, indeed, usually it is not empty and, therefore, we may achieve simultaneous controlled approximate binaural cue preservation and, in most cases, controlled noise reduction of $m > 2M - 2$ interferers. This can be observed experimentally from our results in Sections VI-C2 and VI-C3.

## VI. EXPERIMENTAL RESULTS

In this section, the proposed algorithm is experimentally evaluated. In Section VI-A, the setup of our experiments is demonstrated. In Section VI-B, the performance measures are presented. In Section VI-C, the proposed method is compared to other LCMV-based methods with regard to binaural cue preservation and noise reduction. Moreover, we provide results with regard to the speed of the proposed method in terms of number of iterations.

### A. Experiment Setup

Fig. 2 shows the experimental setup that we used for our experiments. Two behind-the-ear (BTE) hearing aids, with two microphones each, are used for the experiments. Therefore, the total number of microphones is $M = 4$. The publicly available database with the BTE impulse responses (IRs) in [39] is used to simulate the head IRs (we used the front and middle microphone for each hearing aid). The front microphones are selected as reference microphones.

We placed all sources on a $h = 80$ cm radius circle centered at the origin $(0, 0)$ (center of head) with an elevation of $0^o$ degrees. The index of each interferer (denoted by 'x' marker) is indicated in Fig. 2. The interferers $1, 2, 3, 4, 5, 6$ and $7$ are speech shaped noise realizations with the same power and are placed at $15^o, 45^o, 75^o, 105^o, 165^o, 240^o$ and $300^o$ degrees, respectively. The target source (denoted by 'o' marker) is a speech signal in the look direction, i.e., $90^o$ degrees.

The duration of all sources is 60 sec. The microphone self noise at each microphone is simulated as white Gaussian noise (WGN) with $\mathbf{P_V} = \sigma^2 \mathbf{I}$, where $\sigma = 3.8 * 10^{-5}$ which corresponds to an SNR of 50 dB with respect to the target signal at the left reference microphone.

The noise CPSD matrices, $\mathbf{P}$, are calculated (as in Eq. (3)) using the ATFs of the truncated true BTE IRs, from the database, and the estimated PSDs of the sources using all available data without voice activity detection (VAD) errors. Also, the constraints of all the aforementioned methods use the ATFs of the truncated true BTE IRs. The truncated BTE IRs length is 12.5 ms. The sampling frequency is $f_s = 16$ kHz. We use a simple overlap-and-add analysis/synthesis method [40] with frame length 10 ms, overlap 50% and an FFT size of 256. The analysis/synthesis window is a square-root-Hann window. The ATFs are also computed with an FFT size of 256. Finally, the microphone signals are computed by convolving the truncated BTE IRs with the source signals at the original locations.

---

[3]Recall that for $m = 2M - 2$ (i.e., $d = 2M$), there is a feasible solution which does not provide controlled noise reduction (see Section III-B).

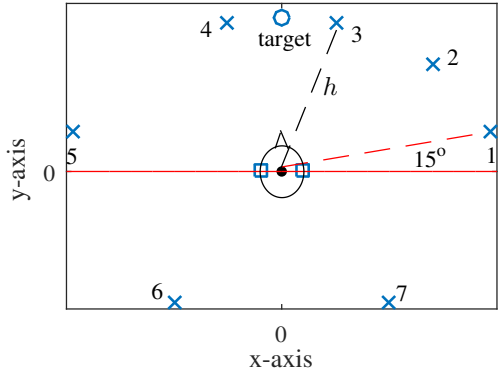

Fig. 2. Experimental setup: □ hearing aids, 'o' target source, 'x' speech shaped interferers. Each source has the same distance, $h$, from the center of the head.

### B. Performance Evaluation

In this section we define the performance evaluation measures that we use to evaluate the results.

*1) ITFs, IPDs & ILDs:* In this section three average performance measures for binaural cue preservation are defined: the average ILD error, the average IPD error, and the average ITF error. Note, as explained in Section III-A, that the IPD errors are perceptually important only for frequencies below 1 kHz, and the ILD errors are perceptually important only for frequencies above 3 kHz. Therefore, the evaluation of IPDs and ILDs will be done only for these frequency regions. We evaluate the average ILD and IPD error for all interferers as follows. Let $\mathcal{L}_{\mathbf{n}_i}(k,l)$ and $\mathcal{T}_{\mathbf{n}_i}(k,l)$ denote the ILD and IPD errors (for the $k$-th frequency bin and $l$-th frame), respectively, defined in Eq. (9). Then the average ILD and ITD errors are defined as

$$\text{TotER}^{\text{ILD}} = \sum_{i=1}^{r}\left(\frac{1}{N-k_{\text{ILD}}}\sum_{k=k_{\text{ILD}}}^{N}\left(\frac{1}{T}\sum_{l=1}^{T}\mathcal{L}_{\mathbf{n}_i}(k,l)\right)\right),$$ (45)

and

$$\text{TotER}^{\text{IPD}} = \sum_{i=1}^{r}\left(\frac{1}{k_{\text{IPD}}}\sum_{k=1}^{k_{\text{IPD}}}\left(\frac{1}{T}\sum_{l=1}^{T}\mathcal{T}_{\mathbf{n}_i}(k,l)\right)\right),$$ (46)

where $N$ and $T$ are the number of frequency bins and the number of frames, respectively, $k_{\text{ILD}}$ and $k_{\text{IPD}}$ are the first and last frequency-bin indices in the frequency regions $3 -$ 8 kHz and $0 - 1$ kHz, respectively.. Note that since the max possible value of $\mathcal{T}_{\mathbf{n}_i}(k,l)$ is 1, the max value of TotER$^{\text{IPD}}$ is $r$. Moreover, we evaluate the average ITF error given by

$$\text{TotER}^{\text{ITF}} = \sum_{i=1}^{r}\left(\frac{1}{N}\sum_{k=1}^{N}\left(\frac{1}{T}\sum_{l=1}^{T}\mathcal{E}_{\mathbf{n}_i}(k,l)\right)\right),$$ (47)

where $\mathcal{E}_{\mathbf{n}_i}$ is the ITF error defined in Eq. (8). Finally, we evaluate the average ITF error ratio given by

$$\text{AvER}^{\text{ITF}}(c) = \frac{1}{r}\sum_{i=1}^{r}\frac{1}{N}\sum_{k=1}^{N}\frac{1}{T}\sum_{l=1}^{T}\frac{\mathcal{E}_{\mathbf{n}_i}(k,l)}{\mathcal{E}_{\mathbf{n}_i,\text{BMVDR}}(k,l)},$$ (48)

which measures the average amount of binaural cue collapse by comparing the ITF error of the proposed method with the ITF error of the BMVDR. Since the proposed method will always satisfy the condition $\mathcal{E}_{\mathbf{n}_i,(k)}(k,l) \leq c\mathcal{E}_{\mathbf{n}_i,\text{BMVDR}}(k,l)$ for $r \leq 2M-3$ (see Theorem 1), obviously AvER$^{\text{ITF}}(c) \leq c$ for $r \leq 2M-3$. Note that ideally the proposed method will provide a solution as close as possible to the boundary of $\Psi(c)$, i.e., AvER$^{\text{ITF}}(c) \approx c$ (see Section V-B). Moreover, for the proposed method AvER$^{\text{ITF}}(0) = 0$ and AvER$^{\text{ITF}}(1) = 1$ because for $c = 0$, $\mathcal{E}_{\mathbf{n}_i}(k,l) = 0$ (for $r \leq 2M-3$), and for $c = 1$, $\mathcal{E}_{\mathbf{n}_i}(k,l) = \mathcal{E}_{\mathbf{n}_i,\text{BMVDR}}(k,l)$.

Please note that in these experiments we use the true ATFs in the constraints of the optimization problems of all competing methods. Therefore, we do not measure the corresponding error measures for the binaural cues of target source since they are always zero, because in all compared methods the constraints perfectly preserve the binaural cues of the target source.

*2) SNR measures:* We define the binaural global segmental signal-to-noise-ratio (gsSNR) gain as

$$\text{gsSNR}^{\text{gain}} = \text{gsSNR}^{\text{out}} - \text{gsSNR}^{\text{in}}.$$ (49)

where the gsSNR input and output are defined as

$$\text{gsSNR}^{\text{in}} = \frac{1}{T}\sum_{l=1}^{T}\min\big(\max\big(\text{SNR}^{\text{in}}(l),-20\big),50\big),$$ (50)

$$\text{gsSNR}^{\text{out}} = \frac{1}{T}\sum_{l=1}^{T}\min\big(\max\big(\text{SNR}^{\text{out}}(l),-20\big),50\big),$$ (51)

respectively, where for the $l$-th frame, the binaural input signal-to-noise-ratio (SNR) is defined as

$$\text{SNR}^{\text{in}}(l) = 10\log_{10}\left(\frac{\sum_{k=1}^{N}\mathbf{e}^T\tilde{\mathbf{P}}_{\mathbf{x}}(k,l)\mathbf{e}}{\sum_{k=1}^{N}\mathbf{e}^T\tilde{\mathbf{P}}(k,l)\mathbf{e}}\right),$$ (52)

where $\mathbf{e}^T = [\mathbf{e}_L^T \quad \mathbf{e}_R^T]$, $\mathbf{e}_L = [1,0,\cdots,0]$ and $\mathbf{e}_R^T = [0,\cdots,0,1]$, $\tilde{\mathbf{P}}$ is defined in Eq. (11) and $\tilde{\mathbf{P}}_{\mathbf{x}}$ is similarly defined but it uses as diagonal block matrices the $\mathbf{P}_{\mathbf{x}}$ matrix. The binaural output SNR for the $l$-th frame, is defined as

$$\text{SNR}^{\text{out}}(l) = 10\log_{10}\left(\frac{\sum_{k=1}^{N}\mathbf{w}^H(k,l)\tilde{\mathbf{P}}_{\mathbf{x}}(k,l)\mathbf{w}(k,l)}{\sum_{k=1}^{N}\mathbf{w}^H(k,l)\tilde{\mathbf{P}}(k,l)\mathbf{w}(k,l)}\right),$$ (53)

where $\mathbf{w} = [\mathbf{w}_L^T(k,l) \quad \mathbf{w}_R^T(k,l)]^T$. Note that gsSNR$^{\text{out}}$ and gsSNR$^{\text{in}}$ can be seen as average measures of the binaural SNR measures defined in [30].

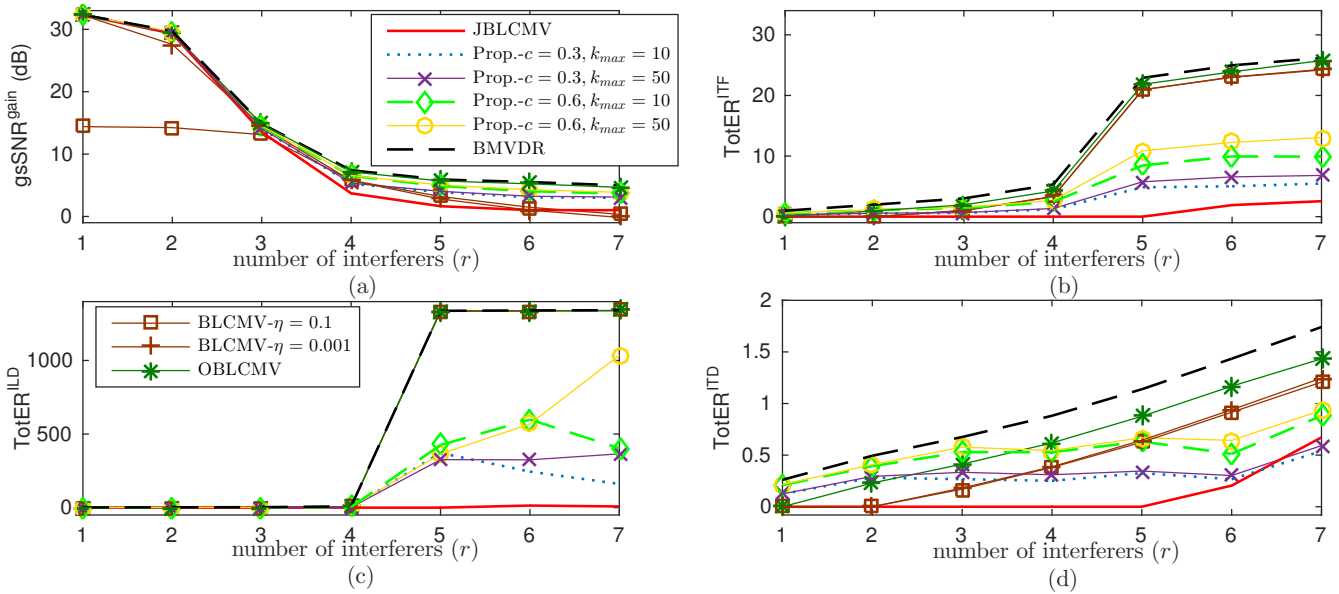

Fig. 3. Anechoic environment: Performance curves for the competing methods in terms of (a) noise reduction, (b) average ITF error, (c) average ILD error, (d) average IPD error.

## C. Results

In the following experiments we evaluate the performance of the proposed and reference methods (i.e., BLCMV [27] with two different values of $\eta$, OBLCMV [28], BMVDR [30] and JBLCMV [29], [30]) as a function of the number of simultaneously present interferers, $1 \le r \le 7$. For instance, for $r = 1$, only the interferer with index 1 is enabled while all the others are silent. For $r = 2$, only the interferers with indices $1, 2$ are enabled, while the others are silent, and so on. The binaural gsSNR$^{in}$ values for $r = 1, 2, 3, 4, 5, 6$ and $7$ are $0.46$, $-1.45$, $-2.29$, $-2.92$, $-3.76$, $-4.15$, and $-4.53$ dB, respectively. Recall that each method has a different $m_{max}$, except for the proposed method for $c > 0$ where $m_{max}$ is difficult to be estimated, as explained in Section V, and, therefore, $m$ is always set to $m = r$. For each of the reference methods and the proposed method in the case of $c = 0$ and if $r > m_{max}$, we will use in the constraints only the first $m_{max}$ interferers and the last $r - m_{max}$ will not be preserved. In Sections VI-C1, VI-C2 the simulations are carried out without taking into account room acoustics. In Section VI-C3 the simulations are carried out by taking into account room acoustics.

### 1) SNR & Binaural Cue Preservation:
For simplicity, we used the same $c = c_j$, for $j = 1, \cdots, m$ for all interferers in the proposed method. In other words, we assumed that the binaural cues of all interferers are equally important. Moreover, we selected for the adaptive change of $\tau_{(k)}$ the step parameter $\alpha(k_{max})$ with $k_{max} \in \{10, 50\}$.

Figs. 3 and 4 show the comparison of the proposed method (denoted by Prop. $-c =$ value, $k_{max} =$ value) with the aforementioned reference methods in terms of binaural cue preservation and noise reduction. Note that BMVDR and the JBLCMV are the two extreme special cases of our method which can be denoted as Prop. $-c = 1$ and Prop. $-c = 0$, respectively. However, in these figures we used the original names for clarity. The performance curves are for different number of simultaneously present interferers $r$. As expected, the performance curves of the proposed method always lie between the BMVDR and the JBLCMV. Fig. 4 is the combination of the curves of Figs. 3(a,b) into a single figure. Notice that the number of interferers $r$ in this combined figure increase from $r = 1$ up to $r = 7$ along the curves from top-left, to bottom-right. As expected, the proposed method for $k_{max} = 50$ achieves slightly better noise reduction and worse binaural cue preservation than for $k_{max} = 10$. This is because for a larger $k_{max}$, the proposed algorithm will provide a feasible solution closer to the boundary of $\Psi(c)$, as explained

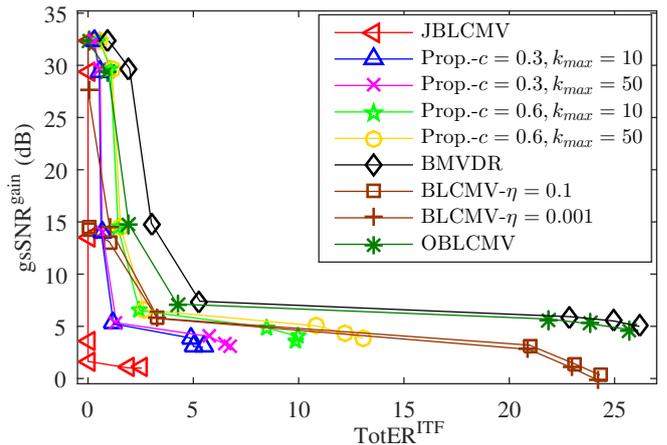

Fig. 4. Anechoic environment: Combination of performance curves from Fig. 3 for the competing methods in terms of (a) noise reduction, (b) average ITF error for different number of simultaneously present interferers $r$. The counting of $r$ starts at the top left part of each curve.

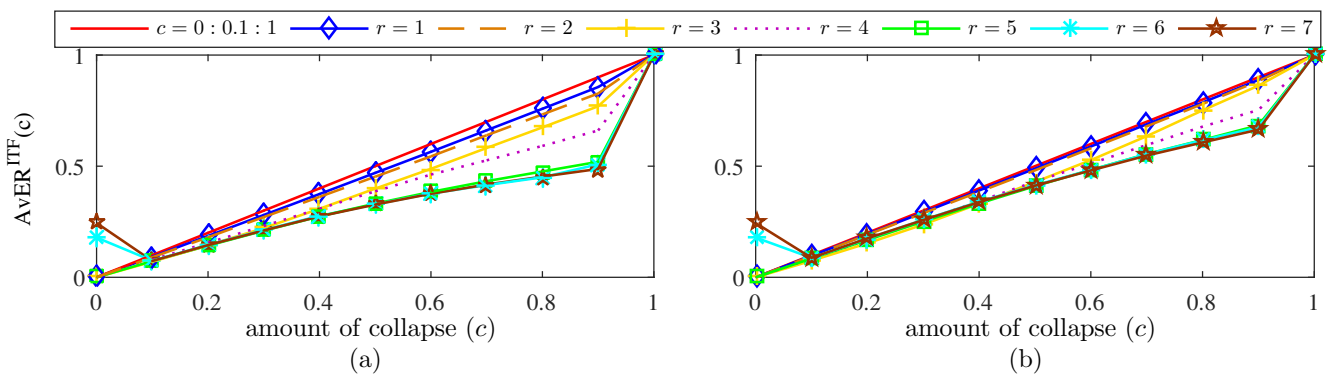

Fig. 5. Anechoic environment: Average ITF error ratio as a function of $c$ for $1 \leq r \leq 7$ for (a) $k_{max} = 10$ and (b) $k_{max} = 50$. The solid line is the $c$ values.

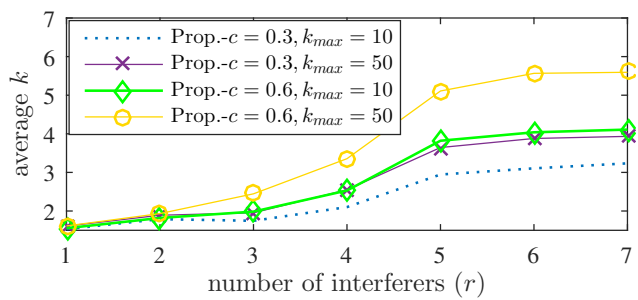

Fig. 6. Anechoic environment: Average number of iterations as a function of simultaneously present interferers, $r$.

in Section V-B.

From Figs. 3(a,b), and Fig. 4 it is clear that, indeed the proposed method achieves a controlled noise reduction and controlled approximate binaural cue preservation. The BMVDR achieves the best noise reduction performance, but it does not preserve the binaural cues of the interferers. The JBLCMV accurately preserves the highest number of simultaneously present interferers and it has worse noise reduction performance than all parametrizations of the proposed method. Note that $m_{max} = 5$ for JBLCMV and, therefore, the last two interferers cannot be included in the constraints and that is why the binaural cue preservation is not perfect. The OBLCMV comes second in terms of SNR performance, but it preserves the binaural cues of only one interferer.

Fig. 5 serves to visualize better the trade-off between fast termination and closeness to the boundary of $\Psi(c)$ (see Section V-B for details). More specifically, Fig. 5 shows the average ITF ratio of the proposed method, for $k_{max} = 10, 50$, as a function of $c$ for different number of simultaneously present interferers $r$. As expected (see Section VI-B1), AvER$^{\text{ITF}}(c) \leq c$ for $1 \leq r \leq 5$. This is also the case for the curves for $r = 6, 7$ except for $c = 0$, as expected, because the proposed method becomes identical to the JBLCMV which can preserve the binaural cue of up to $m_{max} = 2M - 3 = 5$ interferers while achieving controlled noise reduction. As expected, for $k_{max} = 50$ all performance curves are closer to the boundary. In general, the larger the $r$, the less close the AvER$^{\text{ITF}}(c)$ of the proposed method is to $c$. In words, the more constraints we use in the proposed method, the harder it is to reach the boundary of $\Psi(c)$. Note that for the two extreme values $c = 0$ and $c = 1$, the proposed method becomes identical to the JBLCMV and the BMVDR, respectively. As was expected, for $c = 0$ and $r \leq 5$, AvER$^{\text{ITF}}(0) = 0$. The JBLCMV has $m_{max} = 2M - 3 = 5$ and, therefore, for $c = 0$ and $r = 6, 7$, AvER$^{\text{ITF}}(0) > 0$. Finally, for $c = 1$, for all values of $r$, AvER$^{\text{ITF}}(1) = 1$ as expected.

*2) Speed of Termination:* In Fig. 6 we evaluate the number of iterations required for the proposed method to satisfy the stopping criterion (i.e., when it terminates). Fig. 6 shows the average number of iterations as a function of the simultaneously present interferers, $r$, of the four configurations of the proposed method that are tested in Fig. 3 and 4. It is clear that the proposed method terminates after 3-4 iterations on average, even for $r = 6, 7 > 2M - 3$. Note that for both tested values of $k_{max}$, for all frames and frequency bins the proposed method terminated before reaching $k_{max}$.

Fig. 7 shows a 3D histogram which depicts the statistical termination behaviour of the proposed method. Specifically, the proposed method is evaluated with different $c$ values from 0.1 to 0.9 with a step-size 0.1. For each $c$ value it is evaluated for all numbers of simultaneously present interferers, i.e., for $r = 1, \cdots, 7$ as in Fig 6. Hence, this histogram represents all gathered pair-values $(c, k)$ of all frequency bins for all $r = 1, \cdots, 7$. The pairs $(c, k)$ express the number of iterations (per frequency bin), $k$, that the proposed method need in order to terminate for a certain initial $c$. The $z$-axis, which is depicted with different colors, is the number of frequency bins that are associated with a certain pair $(c, k)$ in the x-y axes. Again we see that, on average, after 3-4 iterations the algorithm terminates for $c = 0.1 : 0.1 : 0.9$.

*3) Reverberation:* Figs. 8, 9 and 10 show the same experiments as in Figs. 4, 5, and 6, respectively, but this time in a reverberant office environment. The same signals for the interferers and the target are used here. The BTE IRs containing reverberation are also taken from the database in [39]. Note that, the aforementioned database does not have the reverberant (for the office environment) head IRs corresponding to 240º and 300º degrees [39]. Therefore, we used the avalaible angles, 125º, 145º for the 6-th and 7-th interferer, respectively. Moreover, the sources are now placed on a $h = 100$ cm radius circle centered at the origin $(0, 0)$ (center of head) with an elevation of 0º degrees (because only in this distance is available for the office environment in [39]). The binaural gsSNR$^{in}$ values for $r = 1, 2, 3, 4, 5, 6$ and 7 are now $-0.03$, $-2.1$, $-3.03$, $-3.75$, $-4.44$, $-4.95$, and $-5.42$ dB, respectively.

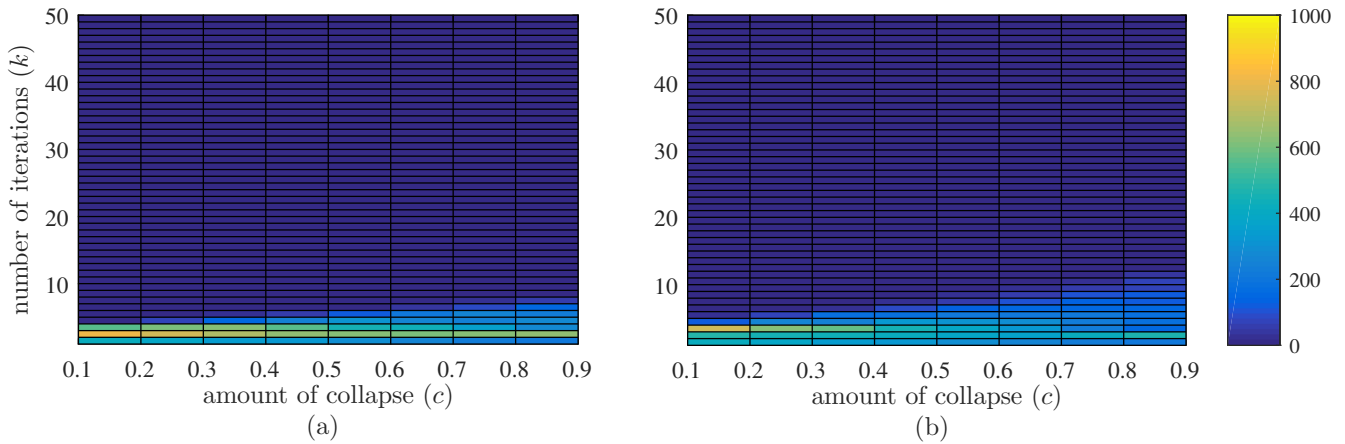

Fig. 7. Anechoic environment: Top view of 3D histogram of number of frequency bins that have pairs $(k, c)$ for the proposed method for (a) $k_{max} = 10$ and (b) $k_{max} = 50$.

As it is shown in Fig. 8, again the performance of the proposed method is bounded between the performance of the BMVDR and the JBLCMV, except for the case with parameters $c = 0.3$ and $k_{max} = 10$ for $r = 7$. The reason for that is that for $r > 2M - 3$ there are no guarantees for bounded performance (see Theorem 1). However, in all other cases of the proposed method for $r > 2M - 3$ the performance is bounded. In Fig. 9 it is clear that the proposed method has very similar behaviour as in Fig. 5, i.e., by increasing $k_{max}$ the proposed method approaches closer to the boundary. Finally, in Fig. 10 it is shown that the speed of termination is not effected significantly due to reverberation.

## VII. Conclusion

In this paper we proposed a new multi-microphone iterative binaural noise reduction method. The proposed method is capable of controlling the amount of noise reduction and the accuracy of binaural cue preservation per interferer using a robust methodology. Specifically, the inequality constraints introduced for the binaural cue preservation of the interferers, are selected in such a way that a) the binaural-cue error is always less or equal than a fraction of the corresponding binaural-cue error of the BMVDR method, and b) the achieved amount of noise suppression is larger or equal to the one achieved via JBLCMV. Therefore, the proposed method provides the flexibility to the users to parametrize the proposed method according to their needs. Moreover, the proposed method always preserves strictly the binaural cues of the target source. Although the proposed method guarantees controlled approximate binaural cue preservation and controlled noise reduction only for $m \leq 2M - 3$ interferers, it is experimentally demonstrated that is also capable of doing the same (in most cases) for more interferers and terminate in just a few iterations.

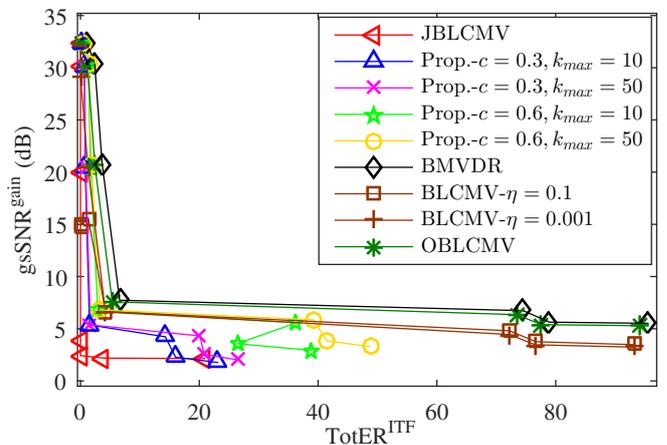

Fig. 8. Reverberant environment (office): Combination of performance curves from Fig. 3 for the competing methods in terms of (a) noise reduction, (b) average ITF error for different number of simultaneously present interferers $r$. The counting of $r$ starts at the top left part of each curve.

## Appendix

In this section, we show how the optimization problem in Eq. (36) can be equivalently written as a second order cone programming (SOCP) problem. For convenience, we reformulate the optimization problem in Eq. (36) using RATFs instead of ATFs. The left and right RATFs of the $i$-th interferer are $\tilde{\mathbf{b}}_{i,L} = (1/b_{iL})\mathbf{b}_i$ and $\tilde{\mathbf{b}}_{i,R} = (1/b_{iR})\mathbf{b}_i$, respectively, while the left and right RATFs of the target are $\bar{\mathbf{a}}_L = (1/a_L)\mathbf{a}$ and $\bar{\mathbf{a}}_R = (1/a_R)\mathbf{a}$, respectively. It is easy to show that the constraints of the optimization problem in Eq. (36) can be

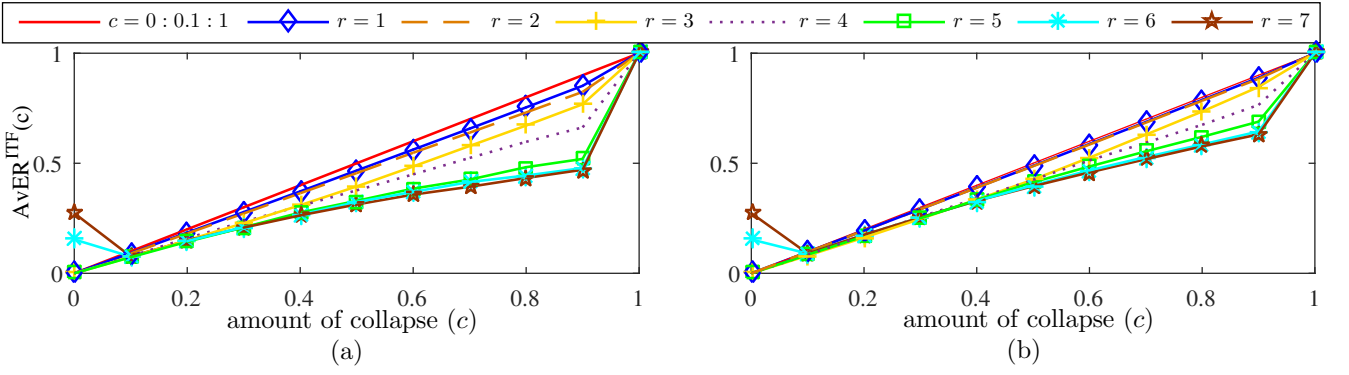

Fig. 9. Reverberant environment (office): Average ITF error ratio as a function of $c$ for $1 \leq r \leq 7$ for (a) $k_{\max} = 10$ and (b) $k_{\max} = 50$. The solid line is the $c$ values.

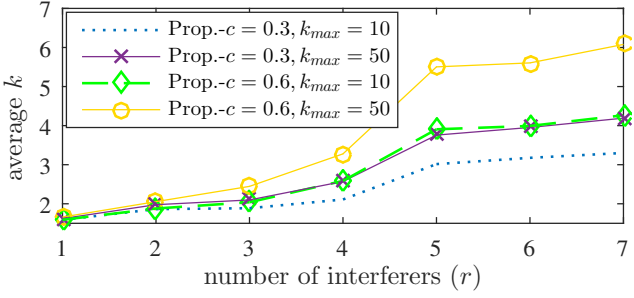

Fig. 10. Reverberant environment (office): Average number of iterations as a function of simultaneously present interferers, $r$.

equivalently written as

$$\underbrace{\begin{bmatrix} \bar{\mathbf{a}}_R^H & \mathbf{0}^H \\ \mathbf{0}^H & \bar{\mathbf{a}}_R^H \end{bmatrix}}_{\boldsymbol{\Phi}_1^H} \mathbf{w} = \underbrace{\begin{bmatrix} 1 \\ 1 \end{bmatrix}}_{\mathbf{q}_1}, \tag{54}$$

$$\left| \boldsymbol{\Phi}_{2,i}^H \mathbf{w} \right| \leq \underbrace{\left| \tau_{(k)} \zeta \bar{\mathbf{b}}_{i,R}^H \hat{\mathbf{w}}_{R,(k-1)} \right|}_{\mathbf{q}_{2,i}}, i = 1, \cdots, m, \tag{55}$$

where $\zeta = |\bar{a}_{R,1}^* \bar{b}_{i,L,M}^* - 1|$ (with $\bar{a}_{R,1}^*$ the first element of $\bar{\mathbf{a}}_R^H$ and $\bar{b}_{i,L,M}^*$ the last element of $\bar{\mathbf{b}}_{i,L}$) and $\boldsymbol{\Phi}_{2,i}$ is the $i$-th column of the matrix $\boldsymbol{\Phi}_2$ given by

$$\boldsymbol{\Phi}_2 = \begin{bmatrix} \bar{\mathbf{b}}_{1L}, \cdots, \bar{\mathbf{b}}_{mL} \\ -\bar{\mathbf{b}}_{1R}, \cdots, -\bar{\mathbf{b}}_{mR} \end{bmatrix}. \tag{56}$$

Similar to [36], [37], we convert the complex vectors and matrices to real-valued ones, i.e.,

$$\breve{\mathbf{w}} = \begin{bmatrix} \breve{\mathbf{w}}_L \\ \breve{\mathbf{w}}_R \end{bmatrix}, \breve{\mathbf{w}}_L = \begin{bmatrix} \mathrm{Re}\{\mathbf{w}_L\} \\ \mathrm{Im}\{\mathbf{w}_L\} \end{bmatrix}, \breve{\mathbf{w}}_R = \begin{bmatrix} \mathrm{Re}\{\mathbf{w}_R\} \\ \mathrm{Im}\{\mathbf{w}_R\} \end{bmatrix}, \tag{57}$$

$$\breve{\mathbf{a}}_L = \begin{bmatrix} \mathrm{Re}\{\bar{\mathbf{a}}_L\} \\ \mathrm{Im}\{\bar{\mathbf{a}}_L\} \end{bmatrix}, \breve{\mathbf{a}}_R = \begin{bmatrix} \mathrm{Re}\{\bar{\mathbf{a}}_R\} \\ \mathrm{Im}\{\bar{\mathbf{a}}_R\} \end{bmatrix} \tag{58}$$

$$\breve{\mathbf{a}}_L = \begin{bmatrix} -\mathrm{Im}\{\bar{\mathbf{a}}_L\} \\ \mathrm{Re}\{\bar{\mathbf{a}}_L\} \end{bmatrix}, \breve{\mathbf{a}}_R = \begin{bmatrix} -\mathrm{Im}\{\bar{\mathbf{a}}_R\} \\ \mathrm{Re}\{\bar{\mathbf{a}}_R\} \end{bmatrix} \tag{59}$$

$$\breve{\mathbf{b}}_{iL} = \begin{bmatrix} \mathrm{Re}\{\bar{\mathbf{b}}_{iL}\} \\ \mathrm{Im}\{\bar{\mathbf{b}}_{iL}\} \end{bmatrix}, \breve{\mathbf{b}}_{iR} = \begin{bmatrix} \mathrm{Re}\{\bar{\mathbf{b}}_{iR}\} \\ \mathrm{Im}\{\bar{\mathbf{b}}_{iR}\} \end{bmatrix}, \tag{60}$$

$$\breve{\mathbf{b}}_{iL} = \begin{bmatrix} -\mathrm{Im}\{\bar{\mathbf{b}}_{iL}\} \\ \mathrm{Re}\{\bar{\mathbf{b}}_{iL}\} \end{bmatrix}, \breve{\mathbf{b}}_{iR} = \begin{bmatrix} -\mathrm{Im}\{\bar{\mathbf{b}}_{iR}\} \\ \mathrm{Re}\{\bar{\mathbf{b}}_{iR}\} \end{bmatrix}, \tag{61}$$

$$\breve{\mathbf{P}} = \begin{bmatrix} \mathrm{Re}\{\mathbf{P}\} & -\mathrm{Im}\{\mathbf{P}\} \\ \mathrm{Im}\{\mathbf{P}\} & \mathrm{Re}\{\mathbf{P}\} \end{bmatrix}, \tilde{\mathbf{P}} = \begin{bmatrix} \breve{\mathbf{P}} & \mathbf{0} \\ \mathbf{0} & \breve{\mathbf{P}} \end{bmatrix}, \tag{62}$$

$$\breve{\boldsymbol{\Phi}}_1 = \begin{bmatrix} \breve{\mathbf{a}}_L & \mathbf{0} & \breve{\mathbf{a}}_L & \mathbf{0} \\ \mathbf{0} & \breve{\mathbf{a}}_R & \mathbf{0} & \breve{\mathbf{a}}_R \end{bmatrix}, \tag{63}$$

$$\breve{\boldsymbol{\Phi}}_2 = \begin{bmatrix} \breve{\mathbf{b}}_{1L}, \cdots, \breve{\mathbf{b}}_{mL} \\ -\breve{\mathbf{b}}_{1R}, \cdots, -\breve{\mathbf{b}}_{mR} \end{bmatrix}, \tilde{\boldsymbol{\Phi}}_2 = \begin{bmatrix} \breve{\mathbf{b}}_{1L}, \cdots, \breve{\mathbf{b}}_{mL} \\ -\breve{\mathbf{b}}_{1R}, \cdots, -\breve{\mathbf{b}}_{mR} \end{bmatrix}. \tag{64}$$

Note that $\mathbf{w}^T \tilde{\mathbf{P}} \mathbf{w} = \|\tilde{\mathbf{P}}^{1/2} \mathbf{w}\|_2^2$, where $\tilde{\mathbf{P}}^{1/2}$ is the principal square root of $\breve{\mathbf{P}}$. The convex optimization problem in Eq. (36) can be equivalently written as

$$\hat{\breve{\mathbf{w}}}_{(k)} = \underset{t, \breve{\mathbf{w}}}{\arg\min} \ t \ \text{s.t.} \ \breve{\mathbf{w}}^T \breve{\boldsymbol{\Phi}}_1 = \breve{\mathbf{q}}_1^T,$$

$$\|\tilde{\mathbf{P}}^{1/2} \breve{\mathbf{w}}\|_2 \leq t$$

$$\left\| \begin{bmatrix} \breve{\boldsymbol{\Phi}}_{2,i}^T \\ \tilde{\boldsymbol{\Phi}}_{2,i}^T \end{bmatrix} \breve{\mathbf{w}} \right\|_2 \leq q_{2,i,(k)}$$

$$\text{for } i = 1, \cdots, m, \tag{65}$$

where $\breve{\mathbf{q}}_1^T = \begin{bmatrix} 1 & 1 & 0 & 0 \end{bmatrix}$, $\breve{\boldsymbol{\Phi}}_{2,i}$ is the $i$-th column of $\breve{\boldsymbol{\Phi}}_2$, and $\tilde{\boldsymbol{\Phi}}_{2,i}$ is the $i$-th column of $\tilde{\boldsymbol{\Phi}}_2$. Note that the problem in Eq. (65) is a standard-form SOCP problem [38].

## Acknowledgment

The authors would like to thank Dr. Meng Guo for his helpful comments and suggestions.